\documentclass{aa}  

\usepackage{graphicx}
\usepackage{txfonts}
\usepackage{siunitx}
\usepackage{tabularx}
\usepackage{multirow}
\usepackage{subcaption}
\usepackage{hyperref}

\newcommand\fermi{{\it Fermi}-LAT}
\newcommand\chandra{{\it Chandra}}
\newcommand\nustar{{\it NuSTAR}}
\newcommand\xmm{XMM-{\it Newton}}
\newcommand\sfr{SFR}
\newcommand\tab{Table~}
\newcommand\fig{Fig.~}
\DeclareSIUnit\gauss{G}
\DeclareSIUnit\parsec{pc}
\DeclareSIUnit\erg{erg}
\DeclareSIUnit\yr{yr}
\DeclareSIUnit\mas{mas}
\DeclareSIUnit\ph{ph}
\DeclareSIUnit\Jy{Jy}

\begin{document}

   \title{Investigating X-ray Emission in the GeV-emitting Compact Symmetric Objects PKS 1718-649 and TXS 1146+596}

   \author{E. Bronzini\inst{1,2}\fnmsep\thanks{\email{\href{mailto:ettore.bronzini@inaf.it}{ettore.bronzini@inaf.it}}}
          \and
          G. Migliori\inst{3}
          \and
          C. Vignali\inst{1,2}
          \and
          M. Sobolewska\inst{4}
          \and
          \L{}. Stawarz\inst{5}
          \and
          A. Siemiginowska\inst{4}
          \and
          M. Orienti\inst{3}
          \and\\
          F. D'Ammando\inst{3}
          \and
          M. Giroletti\inst{3}
          \and
          G. Principe\inst{3,6,7}
          \and
          K. Balasubramaniam\inst{8}
          }

   \institute{Dipartimento di Fisica e Astronomia "Augusto Righi", Università di Bologna, via P. Gobetti 93/2 - 40129 Bologna, Italy
         \and
             INAF, Astrophysics and Space Science Observatory Bologna, via P. Gobetti 93/3 - 40129 Bologna, Italy
         \and
             INAF, Istituto di Radioastronomia, via P. Gobetti 101 - 40129 Bologna, Italy
         \and
            Center for Astrophysics | Harvard \& Smithsonian, 60 Garden Street, Cambridge, MA 02138, USA
        \and
            Astronomical Observatory of the Jagiellonian University, ul. Orla 171, 30-244 Krak\'ow, Poland
        \and
            Dipartimento di Fisica, Università di Trieste, I-34127 Trieste, Italy
        \and
            INFN, Sezione di Trieste, I-34127 Trieste, Italy
        \and
            National Tsing Hua University, Kuang-Fu Road, Hsinchu 30013, Taiwan, R.O.C
         }

   \date{Received XXX; accepted ZZZ}

 
  \abstract
   {}
   {Compact Symmetric Objects (CSOs) are thought to represent the first step in the evolutionary path of radio galaxies. In this work, we investigate the X-ray emission of two CSOs confirmed to emit at GeV energies: PKS 1718-649 and TXS 1146+596. Unveiling the origin of their observed high-energy emission is crucial to establishing the physical parameters of the radio source and understanding how CSOs interact with the surrounding medium.} 
   {We combined archival and new \nustar{} observations of PKS 1718-649 and TXS 1146+596 to have a broadband X-ray coverage. For both sources, we model the broadband spectral energy distribution, from radio band up to $\gamma$-rays, to derive their physical parameters. We also discuss the role of the ambient medium in confining the source expansion, which we investigate using X-ray obscuration.}
   {For the first time, we report on X-ray detections of PKS 1718-649 and 1146+596 with \nustar{} at energies higher than 10 keV. Combining \chandra{} and \nustar{} observations of TXS 1146+596, we reveal the presence of a multi-temperature thermal component dominating the soft X-ray spectrum, and we interpret this finding as indicative of an AGN feedback process in action in this source. In addition, we show that two emitting electrons populations are necessary to reproduce the observed broadband spectral energy distribution of TXS 1146+596: in our models, the X-ray emission could be produced either by synchrotron radiation or by a weak X-ray corona or an ADAF-type emission. Interestingly, an additional X-ray component, i.e. a weak corona, is also required for PKS 1718-649. Moreover, we argue that heavily obscured, and possibly frustrated, sources tend to show different radio sizes with respect to unobscured, free to expand, ones.}
   {}

   \keywords{galaxies: active -- galaxies: jets -- X-rays: galaxies -- radiation mechanisms: non-thermal}
   \titlerunning{Broadband X-ray Emission in GeV-emitting CSOs}
   \maketitle
%


\section{Introduction}
\label{sec: introduction}
Compact Symmetric Objects (hereafter CSOs), defined as sources with radio lobes on both sides of an active nucleus,  which are viewed from the direction perpendicular to the direction of the lobes' expansion, and have an overall size smaller than about one $\si{\kilo \parsec}$, are thought to represent the first step in the evolutionary path of radio galaxies \citep[see][for reviews on CSOs]{Readhead1994,Readhead1996b, Readhead1996a, Readhead2021}. In addition to these two standard criteria, \cite{Kiehlmann2023} recently introduced two additional ones for CSO classification: (i) the fluctuations in flux density, which have to be lower than $\sim 10 \%$ of the total flux density on yearly time-scales, and (ii) the observed speed of components moving along the jets, that have to be less than $2.5\,c$ in any jet component. With this choice, they compiled a sample of 79 \emph{bona-fide} CSOs, including 26 CSOs detected at high-energies (X-rays and/or $\gamma$-rays).

While the CSO classification is based on the sources radio morphology, from the spectral point of view, young radio sources are characterized by a convex peaked synchrotron radio spectrum and are generally split into two classes: Compact Steep Spectrum (CSS) sources having the peak frequency below about 400-500 MHz, and GHz-Peaked Spectrum (GPS) sources having the peak between $\sim 500$ MHz and a few GHz \citep[see][for complete reviews about the radio properties of CSS and GPS]{ODea1997,Orienti2016,ODea2021}. In addition to these, \cite{Dallacasa2000} introduced a further sub-class, the High-Frequency Peakers (HFPs), defined as those with the peak at 5 GHz or even higher in the radio spectrum.

The young nature of CSOs is still a matter of debate. These sources might be compact because they are confined by interaction with dense clouds in the host galaxy interstellar medium (ISM) \citep[e.g.,][]{vanBreugel1984, ODea1991, Dicken2012}, or they might go through a transient or intermittent phase of the radio activity due to e.g. pressure instabilities in accretion disks \citep{Czerny2009}. Jet disruption is another way of producing short-lived radio sources \citep[e.g.,][]{DeYoung1991, Sutherland2007, Wagner2011, Bicknell2018, Mukherjee2018a}, as well as a limited fuel supply to power the jets, for example by a star capture \citep{Readhead1994}. 

Studies of young radio sources at high-energies are useful to constrain important parameters, such as the accretion and jet power, thus shedding a light on their evolution. Furthermore, X-rays are important to determine how these sources release energy in the surrounding medium (if in radiative or mechanical form), hence to study the feedback between CSOs and their host galaxies \citep[see][for a review about the X-ray properties of GPS and CSS sources]{Siemiginowska2009}.  There are several hypotheses for the origin of the X-ray emission in CSOs. It is possible that it is produced primarily by a so-called corona above the accretion disk \citep[e.g.,][]{Tengstrand2009}. However, it is also possible that the X-ray emission is produced in the jets or lobes of the radio source via inverse Compton scattering (IC) off different seed photons \citep[i.e. ultraviolet photons from the accretion disk, infrared photons of the dusty torus, or synchrotron photons from different regions in the jets/lobes;][]{Stawarz2008,Ostorero2010, Migliori2014} and/or by an expanding cocoon that impacts and shocks the ISM of the host galaxy \citep{Ito2011, Kino2013}. 
Due to the milliarcsecond scale linear radio sizes, CSOs remain unresolved by the current X-ray facilities. Therefore, their X-ray morphologies on the scales of their radio lobes cannot be obtained. At the same time, their X-ray spectra are typically described by a power-law model with photon indices in the range $\Gamma \simeq 1.4-1.7$ and luminosities of the order $\mathcal{L}_{2-10 \, \si{\kilo \electronvolt}} \simeq 10^{41}-10^{45} \, \si{\erg \, \second^{-1}}$ \citep[][S19]{Siemiginowska2016,Sobolewska2019b}: these parameters fit well both X-ray corona models and jets/lobes origin, making it difficult to establish the origin of the observed emission.

X-ray observations provide also useful information on the CSOs' environment. The soft band ($\lesssim 2 \, \si{\kilo \electronvolt}$) of their X-ray spectra often reveals the presence of an obscuring medium \citep[$N_{\rm H} \simeq 10^{21} - 10^{24} \, \si{\centi \meter^{-2}}$,][]{Siemiginowska2016}. S19 identified CSOs with moderate-to-high intrinsic hydrogen column densities ($N_{\rm H} > 10^{23} \, \si{\centi \meter^{-2}}$), which, in a linear size versus radio luminosity (at 5 GHz)  plot,  appear to form a branch characterized by smaller sizes in comparison with CSOs having the same 5 GHz luminosity but low intrinsic column densities ($N_{\rm H} \lesssim 10^{22} \, \si{\centi \meter}^{-2}$). This finding suggests that a dense medium in X-ray-obscured CSOs may be able to confine the radio jets, or slow down their expansion. Alternatively, X-ray-obscured CSOs could be brighter in the radio band than their unobscured counterparts because the high-density environments ensure high accretion rates and, possibly, high jet powers. 

Moreover, since the first hard X-ray ($> 10 \, \si{\kilo \electronvolt}$) detection of a CSO \citep[OQ+208,][]{Sobolewska2019a}, it has been clear that broadband X-ray spectra are necessary to fully characterize the properties of CSOs and to probe the medium surrounding the expanding radio source. In particular, \nustar{} observations allow us to extend the X-ray analyses up to a few tens of keV, overcoming issues due to obscuration and better constraining the primary X-ray emission of the sources.

According to theoretical models, the jets and lobes of young radio sources could be the site of production of $\gamma$-ray emission \citep{Stawarz2008,Ostorero2010,Migliori2014}. For CSOs, the model presented in \cite{Stawarz2008} predicts an  isotropic non-thermal emission in the $0.1 - 100 \, \si{\giga \electronvolt}$ band produced in the radio lobes, with luminosities between $10^{41} \, \si{\erg \, \second^{-1}}$ and $ 10^{46} \, \si{\erg \, \second^{-1}}$ depending on the source parameters. 
Alternatively, \cite{Kino2011} discussed a possible hadronic origin for the $\gamma$-ray emission in compact lobes of CSOs. Currently, only a handful of young radio galaxies have been detected in $\gamma$-rays by \fermi{} \citep[see][]{Migliori2016, Principe2020, Lister2020, Principe2021}.
Young radio galaxies are $\gamma$-ray faint, not significantly variable and have $\gamma$-ray photon indices (typically $\Gamma \simeq 2.0-2.6$) and luminosities (generally $\mathcal{L}_{0.1-100 \, \si{\giga \electronvolt}} \simeq 0.9- 1.1 \times 10^{44} \, \si{\erg \, \second^{-1}}$, up to $\mathcal{L}_{0.1-100 \, \si{\giga \electronvolt}} \simeq 1 \times 10^{45} \, \si{\erg \, \second^{-1}}$) similar to misaligned jetted AGN \citep[][]{Principe2021}. Such features are in principle compatible with the origin of the $\gamma$-ray emission (or a fraction of it) in the lobes.

Combining X-ray and $\gamma$-ray information with data from other bands allows us to build and model the Spectral Energy Distribution (SED) of the sources, thus deriving constraints on their physical properties and total jet power. Comparing the inferred results with the most recent numerical simulations of jet-ISM interaction \citep[e.g.,][]{Mukherjee2017} allows us to make predictions on the fate of the sources.
The case of PKS 1718-649 illustrates well the relevance of a $\gamma$-ray detection to our understanding of the origin of the high-energy emission in young radio sources. \cite{Sobolewska2021} performed for the first time modeling of the radio-to-$\gamma$-ray SED of PKS\,1718$-$649. The symmetric morphology rules out significant beamed emission from a jet. Therefore, they applied the dynamic and radiative model of \cite{Stawarz2008} to investigate the emission of the mini-lobes. They showed that, in the framework of this model, the $\gamma$-ray emission is reproduced by IC of the UV photons from an accretion disk off energetic electrons in the mini-radio lobes. Based on the modeling, \cite{Sobolewska2021} could estimate important physical parameters of the source. A moderate departure from energy equipartition conditions, with the particles’ energy density dominating over the magnetic field, is required to reproduce the high-energy emission. The model infers a jet kinetic power of $\mathcal{L}_{\text{kin}} \simeq  2.2 \times 10^{42} \, \si{\erg \, \second^{-1}}$, in agreement with the upper limit $\mathcal{L}_{\text{jet}} \lesssim 2 \times 10^{43} \, \si{\erg \, \second^{-1}}$ reported in \cite{Maccagni2014} based on constrains from Very Long Baseline Interferometry (VLBI)
observations by \cite{Tingay2002}.

The cases of PKS\,1718$-$649 and OQ$+$208 show that a detection in $\gamma$-rays as well as broadening of the observing X-ray window can be strategic to unveiling the nature of high-energy emission in CSOs. In this work, we adopt this approach to study a sample of two CSOs with the goals of (i) investigating the nature of their X-ray and $\gamma$-ray emission, (ii) constraining their physical parameters from their broadband (radio-to-$\gamma$-ray) SED, and (iii) probing the environment in which the sources are expanding.

The paper is organized as follows: in Section \ref{sec: sample selection}, we present the selected sample; then, in Sections \ref{sec: observations and data reduction} and \ref{sec: X-ray analysis}, respectively, we describe the data reduction process and the X-ray analysis performed; in Section \ref{sec: Broadband SED modeling} we present the broadband SED modeling of the sources; we conclude discussing the role of the ambient medium in the radio galaxies evolution and presenting our results about the origin of high-energy emission in CSOs (Section \ref{sec: discussion}).
Throughout this paper, we assume a flat $\Lambda$CDM cosmology and use the most recent estimates of the cosmological parameters \citep{Planck2020}: $\Omega_m = 0.315$, $\Omega_\Lambda = 0.685$ and $H_0 = 67.4 \, \si{\kilo \meter \, \second^{-1} \mega \parsec^{-1}}$.

\section{Sample selection}
\label{sec: sample selection}

For this work, we selected two young radio sources in the sample of \emph{bona-fide} CSOs compiled by \cite{Kiehlmann2023}, which are confirmed to be $\gamma$-ray emitters
\citep{Migliori2016,Principe2020,Fermi_LAT2020b,Principe2021}, with a detection significance $\sigma\geq 5$: PKS 1718-649 (hereafter 1718-649, also known as NGC 6328) and TXS 1146+596 (hereafter 1146+596, also known as NGC 3894). In addition to the archival X-ray data, we obtained \nustar{} observations of the two sources to characterize their X-ray spectrum above $>$10 keV. The basic properties of each source are provided in \tab \ref{tab: sample properties}. In the following we briefly summarize the results of previous multi-wavelength studies.

\begin{table*}
    \centering
    \caption{Main properties of the selected sample.}
    \begin{tabular}{lccccccc}
    \hline
    \hline
    & & & \multicolumn{2}{c}{{\bf 1718-649}} & \multicolumn{2}{c}{{\bf 1146+596}}\\
    \hline
         Parameter & Symbol & Unit & Value & Ref. & Value & Ref.\\
         \hline
         Radio continuum center $\alpha$ (J2000) & RA & & $17^{\si{\hour}} \, 23^{\text{m}} \, 41^{\si{\second}}$ & (1) & $11^{\si{\hour}} \, 48^{\text{m}} \, 36^{\si{\second}}$ & (6)\\
         Radio continuum center $\delta$ (J2000) & DEC & &  $-65{\si{\degree}} \, 00{\si{\arcmin}} \, 37{\si{\arcsecond}}$ & (1) & $-59{\si{\degree}} \, 24{\si{\arcmin}}\, 56{\si{\arcsecond}}$ & (6)\\
         Redshift & $z$ & & 0.014 & (2) & 0.011 & (7) \\
         Kinematic age & $\tau_j$ & yr & 100 & (3) & 60 & (8) \\
         Linear radio size & $LS$ & pc & 2 & (3) & 5 & (8)\\
         Hotspot separation velocity & $v_h$ & $c$ & 0.07 & (3) & 0.20 & (8) \\
         Radio turnover frequency & $\nu_p$ & $\si{\giga \hertz}$ & 3.25 & (4) & 5 & (8)\\
         Galactic column density & $N_{H, \text{Gal}}$ & $\si{\centi \meter^{-2}}/10^{20}$ & $5.90$ & (5) & $1.86$ & (5)\\
         \hline
    \end{tabular}
    \tablefoot{With the assumed cosmology, $1\si{\arcsecond}$ corresponds to $305 \, \si{\parsec}$ at $z =0.014$ and to $228 \, \si{\parsec}$ at $z = 0.011$ \citep{Wright2006}.\\
    {\it References.} (1) \cite{Cutri2003}, (2) \cite{Meyer2004}, (3) \cite{Giroletti2009}, (4) \cite{Tingay2015}, (5) \cite{HI4PI_Collaboration2016}, (6) \cite{deVaucouleurs1991}, (7) \cite{vandenBosch2015}, (8) \cite{Principe2020}.}
    \label{tab: sample properties}
\end{table*}

\subsection{1718-649}
CSO 1718-649, hosted in the galaxy NGC 6328 \citep[$z \simeq 0.014$][]{Meyer2004}, is one of the most studied CSOs. The host galaxy shows a prominent nuclear bulge surrounded by spiral arms of very low surface brightness, which makes its classification uncertain between an S0 or SAB(s)ab galaxy \citep[and references therein]{Maccagni2014}. Spectroscopic observations in the mid-infrared band showed features typical of star-forming gas  \citep[$\sfr \simeq 0.8 - 1.9 \, \mathcal{M}_{\sun} / yr$,][]{Willett2010}.

1718-649 has been extensively studied in the radio band \citep{Bolton1975, Gregory1994,Healey2007,Wright1990,Veron1995,Tingay1997,Tingay2015,Tingay2003,Ricci2006,Sadler2006,Massardi2008,Murphy2010,Maccagni2014,Bennett2003,Chen1995,Massardi2009,Wright2009,Gold2011}
. VLBI observations at 4.8 GHz unveiled two mini-lobes with a linear size $LS\simeq 2 \, \si{\parsec}$ \citep{Tingay1997,Tingay2003, Angioni2019}, while the radio core remains undetected even at $22 \, \si{\giga \hertz}$ \citep{Tingay2003}. The hotspot advance velocity, $\sim 0.07 \, c$ \citep{Giroletti2009}, implies a kinematic age of about $\tau_j \simeq 100$ years.  \citet{Tingay2015} showed that the low-energy radio spectrum is shaped by synchrotron self-absorption and free-free absorption processes. The presence of an inhomogeneous absorbing medium close to the black hole, which could possibly be the gas fuelling the AGN, is also supported by ALMA observations in the sub-mm band \citep{Maccagni2018}. Furthermore, high-quality optical spectra of NGC 6328 obtained by \cite{Filippenko1985} show that high-density ($10^6- 10^7 \, \si{\centi \meter^{-3}}$) clouds exist within $\sim 500 \, \si{\parsec}$ of the nucleus and that photo-ionization is the most favored mechanism responsible for the strong optical emission lines. A weak non-stellar power-law component in the optical spectrum is related to the AGN, classified as a Low Ionization Nuclear Emission-line Region \citep[LINER,][]{Heckman1980}. Estimates of the SMBH mass are of a few 10$^8$ $\mathcal{M}_{\sun}$, based on [O IV] at $25.8 \, \si{\micro \meter}$ transition measurement \citep{Willett2010}.

In X-rays, the source has been first detected with \chandra{} by \cite{Siemiginowska2016}, who identified an unresolved, moderately absorbed, $N_{\rm H} = \left( 0.08\pm0.07 \right) \times 10^{22} \, \si{\centi \meter^{-2}}$, non-thermal ($\Gamma = 1.6\pm0.2$) component, and detected the presence of an extended, diffuse component. Later, combining \chandra{} and \xmm{} observations, \cite{Beuchert2018} refined the spectral parameters of the point-like component ($\Gamma=1.78_{-0.09}^{+0.10}$) and investigated the nature of the extended emission. 
The latter appears to be due to two components: a gas photoionized by the central AGN ($\log \xi = 0.04_{-0.05}^{+1.13}$), and a collisionally ionized plasma ($kT = 0.75^{+0.07}_{-0.08}\, \si{\kilo \electronvolt}$) related to the supernovae activity in the host galaxy. In addition, they reported variability in the intrinsic equivalent hydrogen column density and flux by approximately a factor of 2 over a timescale of 7 years. We note that \cite{Beuchert2018} reported a source-intrinsic column density varying in the range $N_{\rm H} = 0.3 - 0.7 \times 10^{22} \, \si{\centi \meter^{-2}}$, that is about one order of magnitude higher than the values reported in \cite{Siemiginowska2016}: we ascribe this difference to the different choice of the Galactic absorption taken into account. In particular, \cite{Siemiginowska2016} assumed $N_{H,\mathrm{\, Gal}} = 7.15 \times 10^{20} \, \si{\centi \meter^{-2}}$, while \cite{Beuchert2018} $N_{H,\mathrm{\, Gal}} = 5.7 \times 10^{19} \, \si{\centi \meter^{-2}}$. From the latest estimates by \cite{HI4PI_Collaboration2016}, the Galactic attenuation is of the order of $N_{H,\mathrm{\, Gal}} = 5.9 \times 10^{20} \, \si{\centi \meter^{-2}}$, and we then assumed this value for our analyses. 

1718-649 is the first confirmed GPS radio galaxy to be detected in $\gamma$-rays \citep{Migliori2016}. The source is listed in the Fourth \fermi{} catalog of $\gamma$-ray AGN \citep[4LAC,][]{Fermi_LAT2020b}.  
The source is detected at $\sigma\simeq 6$ in about 11 years of \fermi{} data by \cite{Principe2021}: the authors reported a 0.1-100 GeV flux of $\mathcal{F}_{0.1-100 \, \si{\giga \electronvolt}} =\left(  5.30 \pm 1.86 \right) \times 10^{-9}\, \si{\ph \, \centi \meter^{-2} \, \second^{-1}}$ and a corresponding luminosity of $\mathcal{L}_{1-100 \, \si{\giga \electronvolt}} \simeq 1.1 \times 10^{42} \, \si{\erg \, \second^{-1}}$, with a power-law photon index of $\Gamma_{\mathrm{LAT}} = 2.54 \pm 0.17$. No significant $\gamma$-ray variability was found.
The rich radio-to-$\gamma$-ray dataset enabled the first SED modeling, with a particular focus on the high-energy emission \citep{Sobolewska2021}. In the 0.5-10 keV band, the authors concluded that more than one process contributes to the observed X-ray emission: the Comptonization of the IR/torus photons by the electrons in the lobe and a weak additional X-ray component (perhaps an X-ray corona). The authors argue that the two processes should result in a different  $> 10 \, \si{\kilo \electronvolt}$ spectrum. This motivated the \nustar{} observation of 1718-649.

\subsection{1146+596}
\label{sec: 1146+596}
CSO 1146+596 is located in the galaxy NGC 3894 at a redshift $z \simeq 0.011$ \citep{vandenBosch2015}. Multi-epoch VLBI observations unveiled a core-jet parsec-scale morphology \citep{Taylor1998}. Very Long Baseline Array (VLBA) data  at $8.4 \, \si{\giga \hertz}$ presented in \citet{Principe2020} confirmed the compact morphology, with a detected core and two jets at a distance of $5 \, \si{\parsec}$ moving apart at about $0.2 \, c$, seen at an inclination angle with respect to the observer’s line of sight (LOS) of $\si{10\degree < \theta < 21\degree}$. The inferred kinematic age is $\tau_j \simeq 60 \, \text{yrs}$ \citep{Principe2020}. Multi-frequency Very Large Array (VLA) radio observations at 1.4 GHz and 5 GHz of the source pointed out large-scale emission ($\sim 1 \, \si{\kilo \parsec}$) related to multiple radio outbursts \citep{Taylor1998}. 
The estimated mass of the central SMBH is approximately $2 \times 10^9 \, \mathcal{M}_{\sun}$, based on [O IV] at $25 \, \si{\micro \meter}$ transition measurement \citep{Willett2010}, and the AGN is classified as a LINER \citep{Goncalves2004}. NGC 3894 shows an elliptical/lenticular morphology and is classified as a E4-5 galaxy \citep{deVaucouleurs1991}. The optical continuum is dominated by starlight \citep{Condon1988}, and spectroscopic optical observations  \citep{Kim1989,Perlman2001} show the presence of a dust lane and ionized gas along the galaxy’s major axis; the gas kinematics are rather peculiar, exhibiting non-circular motions. 

The first dedicated X-ray study of the system, based on an archival \chandra{} observation, was performed by \cite{Balasubramaniam2021}. As for 1718-649, a point-like source is spatially coincident with the CSO position. 
The core spectrum is best fitted by a combination of a collisionally-ionized thermal plasma with the temperature of $kT = 0.8 \pm 0.1 \, \si{\kilo \electronvolt}$ and a moderately absorbed, hard power-law component ($\Gamma = 1.4 \pm 0.4$, $N_{\rm H} = 2.4 \pm 0.7 \times 10^{22} \, \si{\centi \meter^{-2}}$).  
Interestingly, they also report the detection of the iron K$\alpha$ line at $E_{\ell} = 6.5 \pm 0.1 \, \si{\kilo \electronvolt}$ with a large equivalent width $EW=1.0^{+0.9}_{-0.5}$ keV, possibly indicative of X-ray reflection from a cold neutral absorber located in the central region of the host galaxy.
Typically, $EW$ of this order is expected from Compton-thick AGN (CT AGN) whose primary X-ray emission is absorbed with $N_{\rm H} > 1.5 \times 10^{24}$\,cm$^{-2}$ giving rise to an X-ray reflection-dominated spectrum \citep[e.g.,][and references therein]{Hickox2018}.
Thus, the existing high-quality broadband data allowed \cite{Balasubramaniam2021} to argue that the absorbing column density could be significantly underestimated, making 1146+596 a CT AGN candidate.
An alternative is that the underlying continuum inferred from only the soft X-ray band data could be biased, leading to a biased estimate of the equivalent width of the iron line. We additionally supplemented this dataset by acquiring the \nustar{} X-ray data extending beyond 10 keV to obtain a reliable measure of the intrinsic column density, to establish the true shape of the X-ray continuum, and perform a test of radiative models as in the case of 1718-649 \citep[][]{Sobolewska2021}. Thus, a high-quality broadband dataset exists for 1146+596, and it allowed \cite{Balasubramaniam2021} to estimate the accretion rate ($\lambda_{\text{Edd}} \equiv \mathcal{L}_{\text{bol}}/\mathcal{L}_{\text{Edd}} \sim 10^{-4}$) and the minimum kinetic power of the CSO jets ($\mathcal{L}_{\text{kin}} \simeq 2 \times 10^{42}\, \si{\erg\, \second^{-1}}$).

1146+596 has been detected in $\gamma$-rays by \fermi{} with a significance of $\sigma \simeq 9.7$ and no significant variability has been observed in the $\gamma$-ray flux on a yearly time-scale \citep{Principe2020}. The $\gamma$-ray spectrum is well modeled by a flat power-law ($\Gamma_{\mathrm{LAT}} = 2.05 \pm 0.09$) and a flux of $\mathcal{F}_{0.1-100 \, \si{\giga \electronvolt}} = \left(2.2 \pm 1\right) \times 10^{-9} \, \si{\ph \, \second^{-1} \, \centi \meter^{-2}}$, with a corresponding luminosity of $\mathcal{L}_{1-100 \, \si{\giga \electronvolt}} \simeq 6 \times 10^{41} \, \si{\erg \, \second^{-1}}$. Noteworthy, 1146+596 is the only CT AGN candidate among the young radio galaxies detected in $\gamma$-rays so far.

In view of its multiwavelength features, the sources is an ideal laboratory to investigate at the same time the origin of the high-energy emission in young radio sources and the physics of the feedback process during the early stages of a radio source expansion.

\section{Observations and data reduction}
\label{sec: observations and data reduction}

In the following, we present the observations used in this work. We analyzed the new \nustar{} observations of 1718-649 and 1146+596. We also analyzed the latest \xmm{} observation of 1718-649 for the first time and reanalyzed the \chandra{} observation of 1146+596 presented in \cite{Balasubramaniam2021}. In all of the available datasets, the source extraction region was chosen to maximize the signal-to-noise ratio.

\begin{table*}
	\caption{X-ray observations of 1718-649 and 1146+596 analyzed in this work.}
	\label{tab: sources obsid}
 \small
	\begin{tabular}{cccccccc}
	    \hline
		\hline
		Telescope & Instrument & ObsID. & Date & Nom. exp. time & Fin. exp. time & Ext. reg. & Net cts \\
		& & & & $\left[ \si{\kilo \second}\right]$ & $\left[ \si{\kilo \second}\right]$  & \\
		(1) & (2) & (3) & (4) & (5) & (6) & (7) & (8)\\
		\hline\\[-2.5mm]
        \multicolumn{8}{c}{\bf 1718-649}\\ \\[-2.5mm]
        \hline
        \xmm & pn & 0845110101 & 2020-03-27/29 & 140.7 & 108.0 & $25\si{\arcsecond}$ & $14735 \pm 121$\\
		\nustar & FPMA (FPMB) & 60601020002 & 2020-08-27 & 68.9 & 68.9 (68.3) & $30\si{\arcsecond}$ & $382 \pm 20$ ($380 \pm 20$)\\
        \hline\\[-2.5mm]
        \multicolumn{8}{c}{\bf 1146+596}\\ \\[-2.5mm]
        \hline
		{\it Chandra} & ACIS-S & 10389$^{\left( a \right)}$ & 2009-07-20 & 38.5 & 38.5 & $2.5 \si{\arcsecond}$ ($10 \si{\arcsecond}$) & $376 \pm 19$ ($567 \pm 24$) \\
		{\it NuSTAR} & FPMA (FPMB) & 60601019002 & 2020-08-02/03 & 77.4 & 77.4 (76.7) & $30\si{\arcsecond}$ & $189\pm 14$ ($158 \pm 13$)\\
		\hline
	\end{tabular}
	\tablefoot{(1) Telescope name, (2) instrument name, (3) observation ID, $^{\left( a \right)}$originally presented in \cite{Balasubramaniam2021}, (4) date of the observation, (5) nominal exposure time, (6) final exposure time after cleaning thread was applied, (7) radius of the circular region used to extract the source spectrum, (8) source counts in the extraction region of the source spectrum. Poissonian errors are listed at $1 \, \sigma$.}
\end{table*}

\subsection{\chandra}

We reanalyzed the archival $38.5 \, \si{\kilo \second}$ \chandra{} ACIS-S observation of 1146+596 (ObsID=10389, see \tab \ref{tab: sources obsid} for details) previously reported in \cite{Balasubramaniam2021}. We used \chandra{} CIAO software v.4.13\footnote{\url{https://cxc.cfa.harvard.edu/ciao/}} \citep{Fruscione2006} and CALDB v.4.9.4 for data processing. We reprocessed the data by running the CIAO tool \texttt{chandra\_repro}. 

The data reduction was carried out following the standard steps. The observation was performed in VFAINT mode. We verified that the observation was not affected by periods of high-flares, therefore the exposure time of the final, cleaned file is basically unchanged (see \tab \ref{tab: sources obsid}). We checked the observation did not suffer pile-up issues and then the task \texttt{specextract} was used to obtain source and background spectra. The source spectrum was extracted from a circular region of radius $2.5\si{\arcsecond}$ centered on the source. For the background spectrum, we selected two circular free-from-source regions of radius $30\si{\arcsecond}$ on the same chip as the target. Since the spatial analysis revealed the presence of diffuse emission in the soft X-ray band \citep{Balasubramaniam2021}, the source spectrum was also extracted from a circular region of radius $10\si{\arcsecond}$ centered on the source. Data were grouped at 15 counts per energy bin.

\subsection{\xmm}
We acquired new observations of 1718-649 with \xmm{} (see \tab \ref{tab: sources obsid}). We used SAS v.19.1.0\footnote{\url{https://www.cosmos.esa.int/web/xmm-newton/sas/}} software package to reprocess \xmm{} data. First, we generated event files for each EPIC camera using the task \texttt{emproc} and \texttt{epproc} for the MOS and the pn, respectively. Then, we extracted the source light-curve in the 10-12 keV energy range from the event file of each observation, in order to filter the flaring particle background. We filtered data according to the standard routine: the observation was checked for high background periods and all the periods with a count rate above 0.5 cts/s were excluded (see \tab \ref{tab: sources obsid} for nominal and final exposure times). We checked that the observation did not suffer from pile-up. Then, for all the detectors, we extracted the source spectrum from a circular region of radius $25\si{\arcsecond}$ centered on the source. For the background spectrum, we selected a circular free-from-source region of radius $80\si{\arcsecond}$. We grouped the data at 20 counts per energy bin.

\subsection{\nustar}
We acquired new observations of 1718 and 1146 with \nustar{} \citep{Harrison2013} which are $77.4 \, \si{\kilo \second}$ and $68.9\, \si{\kilo \second}$ long, respectively (ObsID=60601020002 and ObsID=60601019002, see \tab \ref{tab: sources obsid} for details). We used Heasoft v.6.28 and NuSTARDAS v.2.0.0\footnote{\url{https://heasarc.gsfc.nasa.gov/docs/nustar/analysis/}} with CALDB v.20211202\footnote{\url{https://heasarc.gsfc.nasa.gov/docs/heasarc/caldb/nustar/}} for data processing. For both observations, we ran the \texttt{nupipeline} for data calibration and screening. Then, with the \texttt{nuproducts} task we extracted the scientific products of both instruments on-board \nustar{}, including spectra. The source spectra were extracted from circular regions of radius $30\si{\arcsecond}$ centered on the source coordinates. For the background spectra, we selected a free-from-source circular region of radius $90\si{\arcsecond}$ for both telescopes. Data were grouped at 20 and 15 counts per energy bin for 1718-649 and 1146+596, respectively.

\section{X-ray analysis}
\label{sec: X-ray analysis}

CSOs have too compact angular sizes \citep[$\lesssim 0.1 \si{\arcsecond}$, e.g.,][]{Orienti2016} to be spatially resolved by current X-ray facilities. However, given the proximity of our targets, it is possible to study the emission on the scales of the host galaxy and beyond by taking advantage of the sub-arcsecond angular resolution of \chandra{} ($\sim 0.5\si{\arcsecond}$ on-axis). The extended X-ray emission has been previously detected in both 1718-649 \citep{Siemiginowska2016,Beuchert2018} and in 1146+596 \citep{Balasubramaniam2021}.

For the spectral analysis, we used \texttt{XSPEC v.12.11.1} \citep[][]{Arnaud1996} and $\chi^2$ fitting statistics to analyze background-subtracted data of the sources. The Levenberg-Marquardt minimization technique was adopted. All the errors are listed at a 90\% level of confidence for one significant parameter \citep{Avni1976}, unless stated otherwise. To evaluate the statistical significance for one additional model component, the Fisher test \citep{Fisher1922} was adopted. Tested models discussed in the following sections are summarized in \tab{\ref{tab: X-ray models}}.

\begin{figure*}
     \centering
     \includegraphics[width=\textwidth]{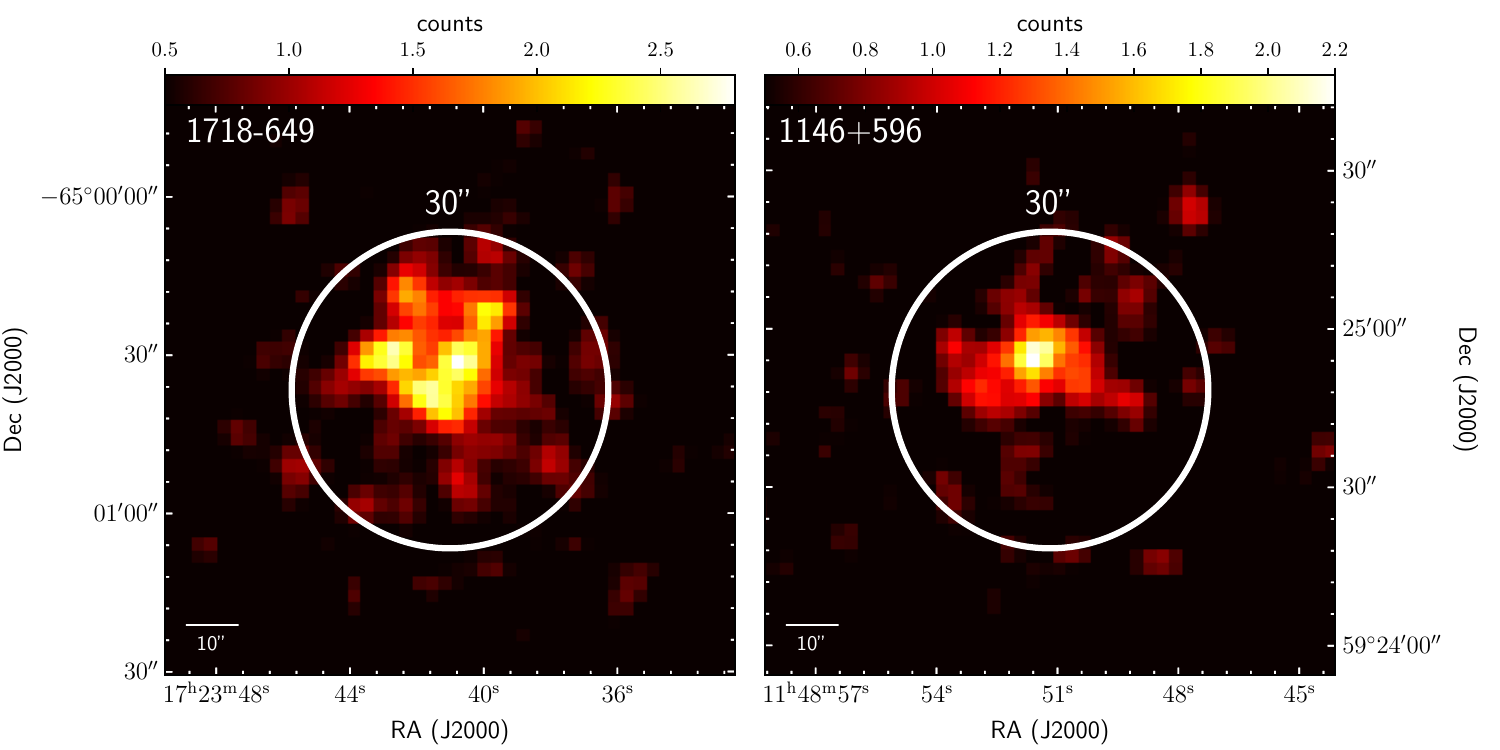}
    \caption{\nustar{}/FPMA images of 1718-649 (\emph{left panel}) and 1146+596 (\emph{right panel}) in the 3-19 keV energy range. The white circular regions of radius $30\si{\arcsecond}$ were used to extract the spectra of the sources. A Gaussian smoothing filter ($\sigma = 1$ pixel, native pixel size) was applied for graphical reasons. The first \nustar{} detection of 1718-649 and 1146+596 are at more than $14 \, \sigma$ and $9 \, \sigma$, respectively.}
    \label{fig: nustar images}
\end{figure*}

\begin{table*}
    \centering
    \caption{Tested models for X-ray analysis.}
    \begin{tabularx}{\textwidth}{cXl}
    \hline
    \hline
         Model & Description & Nomenclature\\
         (1) & (2) & (3)\\
         \hline
         ( I ) & absorbed power-law model including the Galactic absorption & \textsc{phabs$\times$powerlaw};\\
         ( II ) & absorbed power-law model including the Galactic absorption and the intrinsic absorption & \textsc{phabs$\times$zphabs$\times$powerlaw};\\
         ( III ) & absorbed power-law model including the Galactic absorption, the intrinsic absorption, and a thermal plasma & \textsc{phabs$\times$(zphabs$\times$powerlaw+apec)};\\
         ( IV ) & absorbed power-law model including Galactic absorption, the intrinsic absorption, a thermal emitting gas, and a Gaussian emission line & \textsc{phabs$\times$(zphabs$\times$(powerlaw+zgauss)+apec)} ;\\
         ( V ) & absorbed power-law model including Galactic absorption, the intrinsic absorption, two thermal emitting gas, and a Gaussian emission line & \textsc{phabs$\times$(zphabs$\times$(powerlaw+zgauss)+apec+apec)}.\\
         \hline
    \end{tabularx}
    \tablefoot{(1) Model fit number code, (2) description of the model, (3) nomenclature in \texttt{XSPEC}.}
    \label{tab: X-ray models}
\end{table*}

\subsection{1718-649}
We began by modeling the \xmm{} spectra. In the following, only pn detector spectrum is discussed thanks to the higher available statistics with respect to MOS1 and MOS2. However, as a consistency check, MOS data analysis was also performed and the results were consistent with the pn results.

We first assumed a simple power-law model (model I, \textsc{phabs$\times$powerlaw}, in \texttt{XSPEC} nomenclature), where the \textsc{phabs} component accounts for the Galactic absorption in the 0.35-8.5 keV energy range. The best-fit parameters are listed in \tab{\ref{tab: 1718-649 best fit parameters}}, however the reduced $\chi^2$ ($\sim 2$) indicates that this model does not reproduce well the observed spectrum. In fact, the residuals show that the model significantly overestimates the observed data, up to $5 \, \sigma$, below 0.7 keV. Hence, we added a second obscuring component ({\sc zphabs}, in \texttt{Xspec} nomenclature), in order to look for intrinsic absorption of the emission (model II, \textsc{phabs$\times$zphabs$\times$powerlaw}, in \texttt{XSPEC} nomenclature). The fit leaves positive residuals below 2 keV. Given the imaging analysis results indicating the presence of extended emission \citep{Beuchert2018}, we added to the model a thermal component (model III, \textsc{phabs$\times$(zphabs$\times$powerlaw$+$apec)}, in \texttt{XSPEC} nomenclature). The best-fit parameters are listed in \tab \ref{tab: 1718-649 best fit parameters}. The addition of the thermal component improves the fit in a statistically significant way: from the Fisher test, the significance of this addition is larger than $5 \, \sigma$ (see \tab{\ref{tab: 1718-649 best fit parameters}}), confirming the necessity of at least one thermal component ($kT \simeq 0.7 \, \si{\kilo \electronvolt}$) to model the soft band ($<2 \, \si{\kilo \electronvolt}$) of the spectrum, as suggested by \cite{Beuchert2018}. From the fit, we obtained a low level of intrinsic absorption ($N_{\rm H} \simeq 0.8 \times 10^{21} \, \si{\centi \meter^{-2}}$) with a photon index of the power-law component ($\Gamma= 1.73^{+0.05}_{-0.05}$) in agreement with \cite{Beuchert2018} and with the typical values found in CSOs \citep{Siemiginowska2016}. There is no indication of any statistically significant residuals near the energy of the iron lines, also in agreement with the previous studies \citep{Beuchert2018,Siemiginowska2016}.
The source is clearly detected in the 3-19 keV energy range with \nustar{} (see the left panel in \fig{\ref{fig: nustar images}}). The green circle in the figure represents the $30 \si{\arcsecond}$ extraction region of the source spectrum. We performed a simultaneous fit of the FPMA and FPMB spectra (see \tab \ref{tab: 1718-649 best fit parameters}). The best-fit value of the photon index obtained with \nustar{} data is harder ($\Gamma = 2.08_{-0.17}^{+0.17}$) than that from \xmm{} data. However, this tension, about $3 \, \sigma$, appears to be driven by the FPMA instrument ($\Gamma = 2.28_{-0.23}^{+0.24}$), while the fit of the FPMB spectrum  ($\Gamma = 1.86^{+0.24}_{-0.24}$) is in line with the \xmm{} results. Therefore, we conclude that the hardening is likely a consequence of the low-count statistics rather than indicative of a real spectral change. As in the case of \chandra{} data, no statistically significant residuals near the energy of the iron lines were found in \nustar{} data.

Assuming model III, we performed a simultaneous fit between \xmm{}/pn and \nustar{} observations, linking the spectra parameters with the exception of normalizations, for both telescopes. The best-fit parameters are listed in \tab{\ref{tab: 1718-649 best fit parameters}} and the spectra are shown in \fig{\ref{tab: 1718-649 best fit parameters}}. The fit returns a satisfactory value of the reduced $\chi^2$ ($\chi^2/\text{d.o.f.} = 523.4/521$). The fit results are driven by the \xmm{}/pn data. The best-fit value of the photon index of the power-law is $\Gamma=1.76_{-0.05}^{+0.05}$ and the estimated intrinsic luminosity of the source in the 2-10 keV energy range is $\mathcal{L}_{2-10\, \si{\kilo \electronvolt}} = \left(1.59  \pm 0.03 \right) \times 10^{41} \, \si{\erg \, \second^{-1}}$ and the broadband X-ray luminosity of $\mathcal{L}_{0.3-20 \, \si{\kilo \electronvolt}} = \left(3.68  \pm 0.05 \right) \times 10^{41} \, \si{\erg \, \second^{-1}}$, where errors are here listed at $1 \, \sigma$ for one parameter of interest.

\begin{figure}
    \centering
    \includegraphics[width=0.48\textwidth]{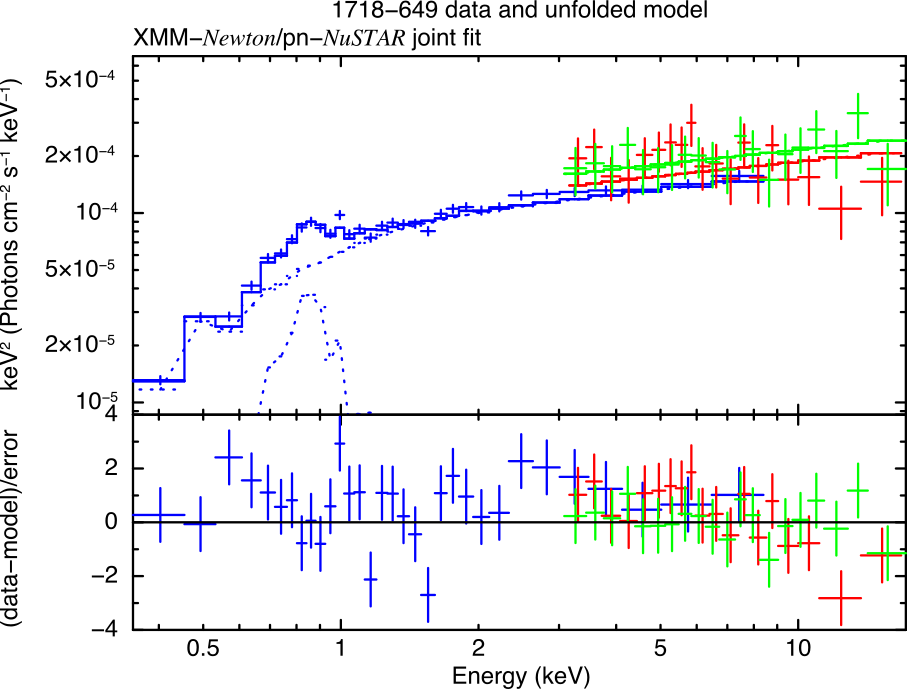}
    \caption{\xmm{}/pn and \nustar{} spectra of 1718-649 fitted with model III in the 0.35-19 keV energy range (see \tab{\ref{tab: 1718-649 best fit parameters}}). \xmm/pn data are shown in blue, while \nustar/FPMA and FPMB data are in red and green, respectively. To make spectra clearer, \xmm{} data are also graphically rebinned.}
    \label{fig: 1718-649 xmm-nustar joint fit}
\end{figure}

Even if the fit returned a satisfactory reduced $\chi^2$ ($\sim 1$), from a visual inspection of \fig{\ref{fig: 1718-649 xmm-nustar joint fit}}, some trends and significant residuals seem to be still present in the 0.35-2 keV energy range, and \cite{Beuchert2018} attempted to mitigate them by including an \textsc{XSTAR} \citep{Bautista2001,Kallman2001} component in their total model. 
Instead, we tested models including emission lines at $\sim 0.6 \, \si{\kilo \electronvolt}$ and $\sim 1 \, \si{\kilo \electronvolt}$, absorption ones at $\sim 1.1 \, \si{\kilo \electronvolt}$ and $\sim 1.5 \, \si{\kilo \electronvolt}$, and a non-equilibrium ionization collisional plasma at $\sim 0.6 \, \si{\kilo \electronvolt}$ (\textsc{nei}, in \texttt{Xspec} nomenclature) instead of the \textsc{apec} component. However, none of these additions is statistically significant to at least $3 \, \sigma$ with a Fisher test. Hence, since we are mainly interested in the X-ray continuum from the CSO, we did not inspect the nature of these possible features in further details.

\begin{table*}
    \centering
    \caption{1718-649 best-fit parameters from X-ray analysis.}
    \begin{tabular}{clccccccc}
         \hline
         \hline
         Model & Abbrv. & \textsc{cons} & \textsc{zphabs} & \multicolumn{2}{c}{\textsc{powerlaw}} & \multicolumn{2}{c}{\textsc{apec}} & $\chi^2/$d.o.f.\\
         & & & $N_H$ & $\Gamma$ & norm & $kT$ & norm\\
         & & & $\left[\times 10^{22} \, \si{\centi \meter^{-2}}\right]$ & & $\left[\times 10^{-4}\right]$ & $\left[\si{\kilo \electronvolt}\right]$ & $\left[\times 10^{-5}\right]$ \\
         (1) & (2) & (3) & (4) & (5) & (6) & (7) & (8) & (9)\\
         \hline \vspace{-0.2cm}\\
         \multicolumn{9}{c}{Individual fits}\vspace{0.1cm}\\
         \hline
         ( I ) & XMM & --- & --- & $1.58_{-0.02}^{+0.02}$ & $0.77_{-0.01}^{+0.01}$ & --- & --- & 850.8/483\\
         ( II ) & XMM & --- & $0.09_{-0.01}^{+0.01}$ & $1.87_{-0.04}^{+0.04}$ & $1.05_{-0.04}^{+0.04}$ & --- & --- & 666.6/482\\
         ( III ) & XMM & --- & $0.08_{-0.02}^{+0.01}$ & $1.73_{-0.05}^{+0.05}$ & $0.88_{-0.05}^{+0.05}$ & $0.72_{-0.05}^{+0.07}$ & $0.96_{-0.12}^{+0.13}$ & 485.6/480\\
         ( I ) & {\it Nu}(A) & --- & --- & $2.28_{-0.23}^{+0.24}$ & $2.96_{-1.06}^{+1.59}$ & --- & --- & 11.9/17\\
         ( I ) & {\it Nu}(B) & --- & --- & $1.86_{-0.24}^{+0.24}$ & $1.51_{-0.57}^{+0.88}$ & --- & --- & 7.6/18\\
         \hline \vspace{-0.2cm}\\
         \multicolumn{9}{c}{Simultaneous fits}\vspace{0.1cm}\\
         \hline
         \multirow{2}{*}{( I )} & {\it Nu}(A) & $1.00$ & \multirow{2}{*}{---} & \multirow{2}{*}{$2.08_{-0.17}^{+0.17}$} & \multirow{2}{*}{$2.06_{-0.60}^{+0.82}$} & \multirow{2}{*}{---} & \multirow{2}{*}{---} & \multirow{2}{*}{23.6/36}\\
         & {\it Nu}(B) & $1.10_{-0.15}^{+0.17}$\vspace{3mm}\\
         
         \multirow{3}{*}{( III )} & XMM & $1.00$ & \multirow{3}{*}{$0.09_{-0.01}^{+0.01}$} & \multirow{3}{*}{$1.76_{-0.05}^{+0.05}$} & \multirow{3}{*}{$0.90_{-0.05}^{+0.05}$} & \multirow{3}{*}{$0.72_{-0.06}^{+0.06}$} & \multirow{3}{*}{$0.93_{-0.12}^{+0.12}$} & \multirow{3}{*}{523.4/521}\\
         & {\it Nu}(A) & $1.19_{-0.13}^{+0.14}$\\
         & {\it Nu}(B) & $1.38_{-0.14}^{+0.15}$\\
         \hline
    \end{tabular}
    \tablefoot{(1) Model (see Table \ref{tab: X-ray models}), (2) abbreviation of the telescope name: XMM stands for \xmm, while {\it Nu} for \nustar{} (FPMA and FPMB are indicated in parentheses), respectively. (3) cross-calibration constant, (4) intrinsic hydrogen column density, (5-6) photon index and normalization of the power law emission component, (7-8) temperature and normalization of the collisionally ionized thermal plasma, (9) ratio of the $\chi^2$ fitting statistics to degrees of freedom. For clarity we split analyses that were performed individually (top panel), from those in which the data taken with different instruments were fitted simultaneously (bottom panel). The models include Galactic hydrogen column density $N_{H, \text{Gal}} = 5.90\times 10^{20} \, \si{\centi \meter^{-2}}$ and thermal component assumes solar abundance. Values without errors were fixed during the fit. The normalizations are $\si{\ph \,\kilo \electronvolt^{-1} \, \centi \meter^{-2} \, \second^{-1}}$ at 1 keV, for the power law component, $10^{-14} (1 + z)^2 n_e n_H V/4 \pi d_L^2$, assuming uniform ionized plasma with electron and $H$ number densities $n_e$ and $n_H$, respectively, and volume $V$, all in cgs units, for the thermal component, and total number of $\si{\ph\, \centi \meter^{-2} \, \second^{-1}}$ in the line, for the Gaussian component.}
    \label{tab: 1718-649 best fit parameters}
\end{table*}

\subsection{1146+596}

Given the presence of extended X-ray emission in the source \citep{Balasubramaniam2021}, we first performed the spectral analysis of the \chandra{} data extracted for the central point-like source within $2.5\si{\arcsecond}$; then, we analyzed the one within a circular region of radius $10\si{\arcsecond}$, which includes the extended emission.

We proceeded testing models of increasing complexity. Models accounting only for absorption (models I and II) have revealed to be too simple to model the complex X-ray spectrum of the source from the $2.5\si{\arcsecond}$ region: the fits returned reduced $\chi^2$ of about 2.7 and 2.9, respectively, and left statistically significant positive residuals (at more than $2 \, \sigma$) below 2 keV. This is not surprising, given the results of the imaging analysis unveiling diffuse emission, and motivated adding a thermal component to the model (model III). This model returns a temperature $kT \simeq 0.84_{-0.16}^{+0.17} \, \si{\kilo \electronvolt}$ for the thermal emission \citep[in agreement with the value $kT = 0.8^{+0.1}_{-0.1} \, \si{\kilo \electronvolt}$ reported in][]{Balasubramaniam2021}. The photon index of the power-law is, however, still unusually hard $(\Gamma = 1.05^{+0.67}_{-0.59})$, in comparison with typical values expected for CSOs \citep{Siemiginowska2016}, possibly indicating that the model does not yet correctly account for the intrinsic absorption. As a test, we fixed $\Gamma = 1.70$, a typical value of unabsorbed CSO \citep[$N_{\rm H} \lesssim 10^{22} \, \si{\centi \meter}^{-2}$;][]{Siemiginowska2016}, obtaining a slight increase, within a factor of 2, of the intrinsic absorption column.
The observed spectrum still displays an excess above the model in the 6-7 keV range. We thus included a Gaussian line at $\sim 6.4 \, \si{\kilo \electronvolt}$ in the model (model IV). 
However, even if model IV better reproduces the spectrum in the 6-7 keV, the inclusion of the Gaussian component does not improve the fit significantly ($1.2\, \sigma$ according to the Fisher test).
The iron line was first reported in \cite{Balasubramaniam2021}, where the line parameters were constrained, $E_\ell = 6.5_{-0.1}^{+0.1} \, \si{\kilo \electronvolt}$ with $\sigma_\ell = 0.12_{-0.08}^{+0.08} \, \si{\kilo \electronvolt}$, without fixing the value of the photon index ($\Gamma = 1.4_{-0.4}^{+0.4}$). This is likely due to a different choice of and count binning (5 counts per bin).

\begin{figure*}
     \centering
     \begin{subfigure}[]{0.45\textwidth}
         \centering
         \includegraphics[width=\textwidth]{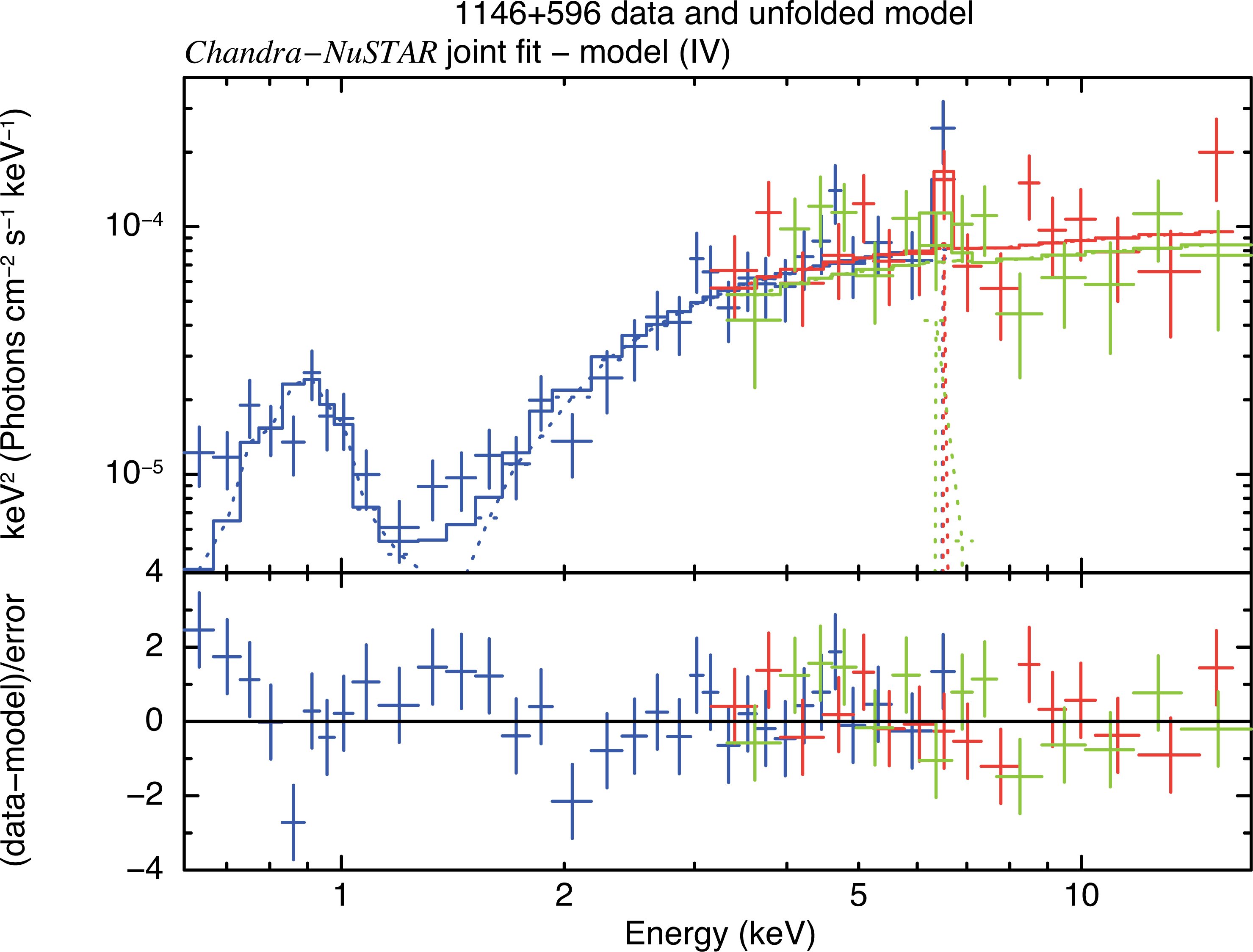}
         \caption{$\chi^2/\text{d.o.f.}=66.7/54$}
         \label{fig: 1146+596 chandra-nustar model 4}
     \end{subfigure}
     \hfill
     \begin{subfigure}[]{0.45\textwidth}
         \centering
         \includegraphics[width=\textwidth]{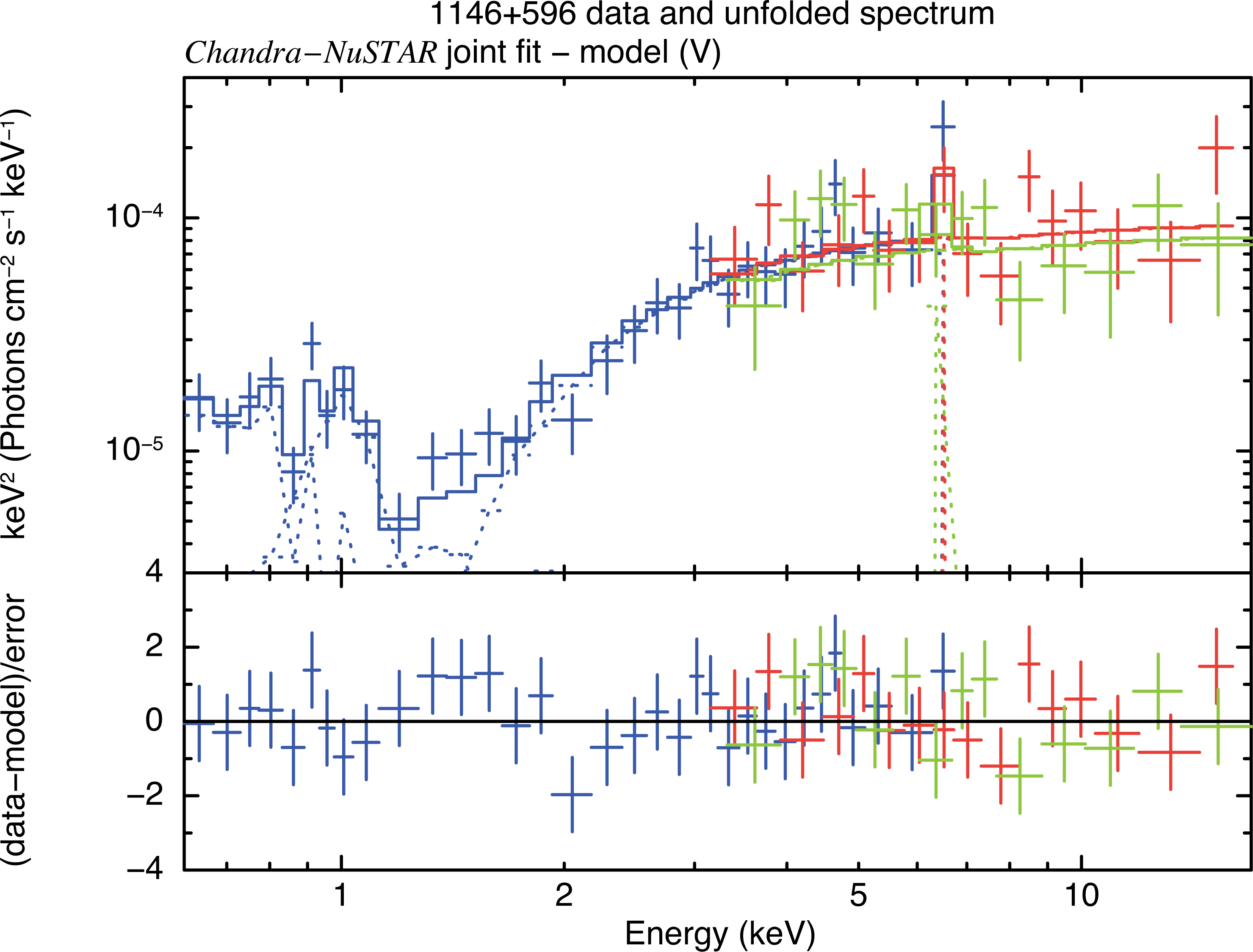}
         \caption{$\chi^2/\text{d.o.f.}=49.1/52$}
         \label{fig: 1146+596 chandra-nustar model 5}
     \end{subfigure}
    \caption{\chandra{} and \nustar{} spectra of 1146+596 fitted with model IV (panel \ref{fig: 1146+596 chandra-nustar model 4}) and model V (panel \ref{fig: 1146+596 chandra-nustar model 5}) in the $0.3-19 \, \si{\kilo \electronvolt}$ energy range (see \tab{\ref{tab: 1146+596 best fit parameters}}). \chandra{}, \nustar{} FPMA, and \nustar{} FPMB data are shown in blue, red, and green, respectively.}
\end{figure*}

Next, we investigated the physical properties of the diffuse emission, by modeling the spectrum extracted from the $10\si{\arcsecond}$ radius region. We started modeling data from an absorbed power-law model and a Gaussian line plus a thermal component (model IV), leaving all the parameters free to vary. However, the fit returned an unusually hard photon index, as in the case of point-like study. Following the same arguments discussed above, we fixed it to 1.70. The best-fit parameters are listed in \tab \ref{tab: 1146+596 best fit parameters}. The results obtained are consistent with those of the point-like emission. However, from the fit, we noticed a relevant scatter in the residuals up to $3\, \sigma$ at energies below 2 keV (see \fig \ref{fig: 1146+596 chandra-nustar model 4}). Hence, we tested a model with a multi-temperature gas (model V) and fit the data with the model \textsc{phabs$\times$(zphabs$\times$(powerlaw$+$zgauss)$+$apec$+$apec)}, in \texttt{XSPEC} nomenclature. The best-fit parameters are listed in \tab\ref{tab: 1146+596 best fit parameters}. The temperature values of the two thermal components are $kT = 0.31^{+0.16}_{-0.09} \, \si{\kilo \electronvolt}$ and $kT = 1.12^{+0.25}_{-0.17} \, \si{\kilo \electronvolt}$. 

The thermal component needed to model the point-like spectrum has an intermediate value $kT \simeq 0.9 \, \si{\kilo \electronvolt}$. This could be explained with a lower number of counts in the $2.5\si{\arcsecond}$ region compared with the larger region, which did not allow us to detect the two-temperature plasma.

The source is clearly detected in the 3-19 keV energy range with \nustar{} (see the right panel in \fig{\ref{fig: nustar images}}). The green circle represents the $30\si{\arcsecond}$ extraction region of the source spectrum. We fit the spectra of the two \nustar{} modules both individually and simultaneously, and no statistically significant discrepancies have been found between the two detectors. The best-fit parameters are listed in \tab\ref{tab: 1146+596 best fit parameters}. From the simultaneous fit, the spectrum is well fitted by a simple power-law model (model I) with a photon index of $\Gamma = 1.91_{-0.27}^{+0.27}$ and no significant signatures of emission lines, including the iron K$\alpha$ line, are present in the spectrum. However, the source net counts in the \chandra{} observation are twice the photon statistics of \nustar{} for each instrument.

Assuming model V, we performed a simultaneous fit of the \chandra{}\footnote{The \chandra{} spectrum is the one extracted from a circular region of radius $10\si{\arcsecond}$ centered on the source.} and \nustar{} observations. The best-fit parameters are listed in \tab\ref{tab: 1146+596 best fit parameters} and the data and the unfolded spectrum are shown in \fig \ref{fig: 1146+596 chandra-nustar model 5}. The simultaneous fit constrains the intrinsic power-law shape:  the photon index is $\Gamma = 1.92^{+0.34}_{-0.33}$ with an intrinsic hydrogen column density of $N_{\rm H} = 3.49^{+1.28}_{-1.04} \times 10^{22} \, \si{\centi \meter^{-2}}$. We also confirm the need for a multi-temperature thermal component, with $kT = 0.32_{-0.09}^{+0.17}$ and $kT = 1.16_{-0.18}^{+0.31}$. For comparison, in \fig{\ref{fig: 1146+596 chandra-nustar model 4}} we show the source spectrum fitted with a single thermal component (model IV), which displays the presence of significant residuals below 2 keV.
From the fit, we estimated the intrinsic luminosity of the source in the 2-10 keV energy range, $\mathcal{L}_{2-10 \, \si{\kilo \electronvolt}} = \left( 6.0 \pm 0.4 \right) \times 10^{40} \, \si{\erg \, \second^{-1}}$, and an unabsorbed broadband X-ray luminosity $\mathcal{L}_{0.3-20 \, \si{\kilo \electronvolt}} = \left(1.45 \pm 0.14 \right) \times 10^{41} \, \si{\erg \, \second^{-1}}$, where errors are listed at the $1 \, \sigma$ confidence level for one parameter of interest.

\begin{table*}
    \centering
    \scriptsize
    \caption{1146+596 best-fit parameters from X-ray analysis.}
    \begin{tabular}{clcccccccccccccc}
         \hline
         \hline
         Model & Abbrv. & \textsc{cons} & \textsc{zphabs} & \multicolumn{2}{c}{\textsc{powerlaw}} & \multicolumn{2}{c}{\textsc{zgauss}} & \multicolumn{2}{c}{\textsc{apec}} & \multicolumn{2}{c}{\textsc{apec}} & $\chi^2/$d.o.f.\\
         & & & $N_H$ & $\Gamma$ & norm & $E_\ell \left(EW \right)$ & norm & $kT$ & norm & $kT$ & norm\\
         & & & $\left[\times 10^{22} \, \si{\centi \meter^{-2}}\right]$ & & $\left[\times 10^{-4}\right]$ & $\left[\si{\kilo \electronvolt }\right]$ & $\left[\times 10^{-6}\right]$ & $\left[\si{\kilo \electronvolt}\right]$ & $\left[\times 10^{-5}\right]$ & $\left[\si{\kilo \electronvolt}\right]$ & $\left[\times 10^{-5}\right]$\\
         (1) & (2) & (3) & (4) & (5) & (6) & (7) & (8) & (9) & (10) & (11) & (12) & (13)\\
         \hline \vspace{-0.2cm}\\
         \multicolumn{13}{c}{Individual fits}\vspace{0.1cm}\\
         \hline
         ( III ) & {\it Ch}$\, \left(2.5\si{\arcsecond}\right)$ & --- &  $1.90_{-1.00}^{+1.32}$ & $1.05_{-0.59}^{+0.67}$ & $0.19_{-0.11}^{+0.34}$ & --- & --- & $0.84_{-0.16}^{+0.17}$ & $0.27_{-0.07}^{+0.08}$ & --- & --- & 31.6/18\\
         ( III ) & {\it Ch}$\, \left(2.5\si{\arcsecond}\right)$ & --- &  $3.04_{-0.55}^{+0.68}$ & $1.70$ & $0.50_{-0.08}^{+0.08}$ & --- & --- & $0.88_{-0.15}^{+0.17}$ & $0.29_{-0.07}^{+0.08}$ & --- & --- & 34.1/19\\
         ( IV ) & {\it Ch}$\, \left(2.5\si{\arcsecond}\right)$ & --- &  $2.93_{-0.56}^{+0.67}$ & $1.70$ & $0.48_{-0.08}^{+0.08}$ & $6.42_{-0.09}^{+0.69} \left( 0.9_{-0.8}^{+0.8} \right)$ & $1.86_{-1.37}^{+1.60}$ & $0.88_{-0.15}^{+0.16}$ & $0.29_{-0.07}^{+0.08}$ & --- & --- & 28.6/17\\
         ( IV ) & {\it Ch}$\, \left(10\si{\arcsecond}\right)$ & --- &  $2.70_{-0.52}^{+0.74}$ & $1.70$ & $0.50_{-0.08}^{+0.08}$ & $6.52_{-0.99}^{+3.75} \left( 0.7_{-0.6}^{+1.2} \right)$ & $1.50_{-0.11}^{+0.98}$ & $0.84_{-0.13}^{+0.11}$ & $0.60_{-0.09}^{+0.11}$ & --- & --- & 36.8/27\\
         ( V ) & {\it Ch}$\, \left(10\si{\arcsecond}\right)$ & --- &  $2.93_{-0.63}^{+0.77}$ & $1.70$ & $0.52_{-0.08}^{+0.08}$ & $6.54_{-1.66}^{+0.46} \left( <3.4 \right)$ & $1.49_{-1.15}^{+1.65}$ & $0.31_{-0.09}^{+0.16}$ & $0.61_{-0.25}^{+0.27}$ & $1.12_{-0.17}^{+0.25}$ & $0.52_{-0.17}^{+0.23}$ & 19.6/25\\
         ( I ) & {\it Nu}(A) & --- & --- & $1.83_{-0.36}^{+0.37}$ & $0.60_{-0.31}^{+0.59}$ & --- & --- & --- & --- & --- & --- & 12.2/14 \\
         ( I ) & {\it Nu}(B) & --- & --- & $2.01_{-0.41}^{+0.41}$ & $0.76_{-0.43}^{+0.86}$ & --- & --- & --- & --- & --- & --- & 14.8/12 \\
         \hline \vspace{-0.2cm}\\
         \multicolumn{13}{c}{Simultaneous fits}\vspace{0.1cm}\\
         \hline
         \multirow{2}{*}{( I )} & {\it Nu}(A) & $1.00$ & \multirow{2}{*}{---} & \multirow{2}{*}{$1.91_{-0.27}^{+0.27}$} & \multirow{2}{*}{$0.70_{-0.29}^{+0.47}$} & \multirow{2}{*}{---} & \multirow{2}{*}{---} & \multirow{2}{*}{---} & \multirow{2}{*}{---} & \multirow{2}{*}{---} & \multirow{2}{*}{---} & \multirow{2}{*}{27.3/27}\\
         & {\it Nu}(B) & $0.90_{-0.18}^{+0.22}$ &\vspace{0.3cm}\\
         
         \multirow{3}{*}{( V )} & {\it Ch}$\, \left(10\si{\arcsecond}\right)$ & $1.00$ & \multirow{3}{*}{$3.49^{+1.28}_{-1.04}$} & \multirow{3}{*}{$1.92_{-0.33}^{+0.34}$} & \multirow{3}{*}{$0.74_{-0.32}^{+0.57}$} & \multirow{3}{*}{$6.73_{-0.35}^{+0.24} \left(0.3^{+0.2}_{-0.2}\right)$} & \multirow{3}{*}{$0.81_{-0.66}^{+0.88}$} & \multirow{3}{*}{$0.32_{-0.09}^{+0.17}$} & \multirow{3}{*}{$0.61_{-0.25}^{+0.27}$} & \multirow{3}{*}{$1.16_{-0.18}^{+0.31}$} & \multirow{3}{*}{$0.56_{-0.19}^{+0.30}$} & \multirow{3}{*}{$49.1/52$}\\
        & {\it Nu}(A) & $1.01^{+0.29}_{-0.23}$\\
         & {\it Nu}(B) & $0.90^{+0.28}_{-0.22}$ \\
         \hline
    \end{tabular}
    \tablefoot{(1) Model (see Table \ref{tab: X-ray models}), (2) abbreviation of the telescope name: {\it Ch} stands for \chandra, while {\it Nu} for \nustar{} (FPMA and FPMB are indicated in parentheses), respectively. For \chandra{} observations the radius of the source extraction region is specified in parentheses. (3) cross-calibration constant, (4) intrinsic hydrogen column density, (5-6) photon index and normalization of the power law emission component, (7-8) energy and normalization of the Gaussian component, where the corrisponding equivalent width $EW$ is reported in parentheses, (9-10) temperature and normalization of the collisionally ionized thermal plasma, (11-12) temperature and normalization of the second collisionally ionized thermal plasma, (13) ratio of the $\chi^2$ fitting statistics to degrees of freedom. For clarity we split analyses that were performed individually (top panel), from those in which the data taken with different instruments were fitted simultaneously (bottom panel). The models include Galactic hydrogen column density $N_{H, \text{Gal}} = 1.86\times 10^{20} \, \si{\centi \meter^{-2}}$ and thermal component assumes solar abundance. For the Gaussian component, the $\sigma$ was set fixed to 10 eV. Values without errors were set fixed during the fit. The normalizations are $\si{\ph \,\kilo \electronvolt^{-1} \, \centi \meter^{-2} \, \second^{-1}}$ at 1 keV, for the power law component, $10^{-14} (1 + z)^2 n_e n_H V/4 \pi d_L^2$, assuming uniform ionized plasma with electron and $H$ number densities $n_e$ and $n_H$, respectively, and volume $V$, all in cgs units, for the thermal component, and total number of $\si{\ph\, \centi \meter^{-2} \, \second^{-1}}$ in the line, for the Gaussian component.}
    \label{tab: 1146+596 best fit parameters}
\end{table*}
\section{Broadband SED modeling}
\label{sec: Broadband SED modeling}

From the X-ray analysis (Section \ref{sec: X-ray analysis}), we obtained a good characterization of the X-ray continuum for both sources. The SED modeling of 1718-649 was presented in \cite{Sobolewska2021} and here we will discuss it in relation to the broadband X-ray results, while we will focus on the broadband SED modeling of 1146+596. Note that in the SED of each source we took into account only the X-ray flux of the power-law component, corrected for absorption, given that our goal was to investigate the origin of the high-energy emission of the radio source.

\subsection{1718-649 SED}
\label{sec: SED 1718-649}

\begin{figure}
    \centering
    \includegraphics[width=0.48\textwidth]{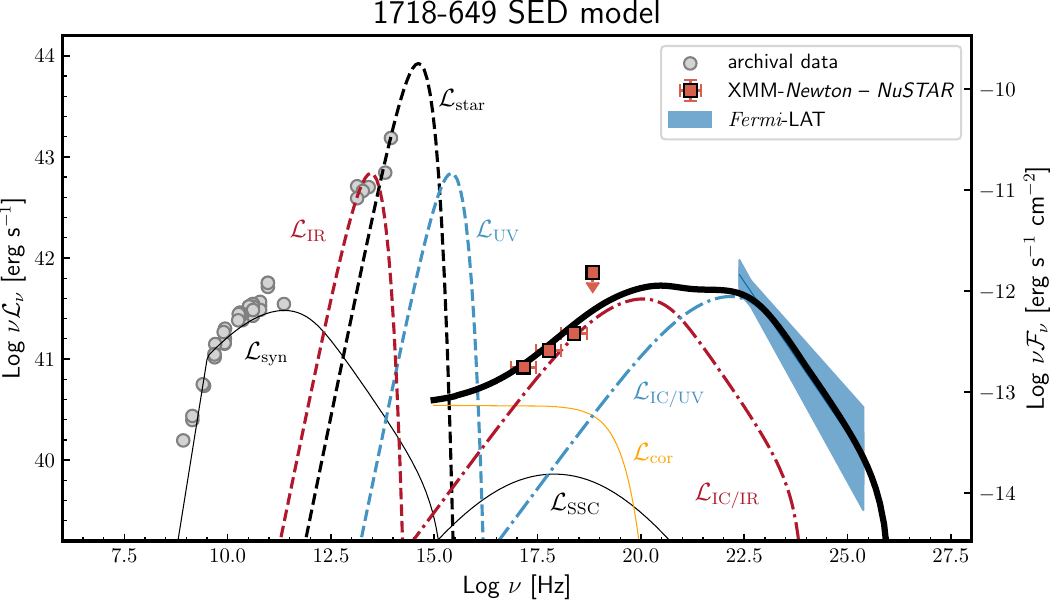}
    \caption{Observed broadband spectral energy distribution of 1718-649 and theoretical model by \protect\cite{Sobolewska2021}. No formal fitting was performed. Gray circles are archival data from \protect\cite{Sobolewska2021} and references therein. The butterfly light blue region is \fermi{} constraints from \protect\cite{Principe2021}. Dark orange data are \xmm-\nustar{} luminosity constrains from this work. 
    Broadband model components are as follows: self-absorbed synchrotron radiation and SSC (the solid thin black line); dashed lines represent three blackbody components (IR in dark red, starlight in black, accretion disk photon fields in light blue), and their corresponding IC components originating from a single radio lobe with the same colors. The solid orange line illustrates the contribution of a low-luminosity X-ray nuclear emission (a weak X-ray corona or an ADAF-type emission). The black solid line is the sum of the different high-energy components.}
    \label{fig: 1718-649 SED}
\end{figure}

The SED modeling presented in \cite{Sobolewska2021} leaves the question on the origin of the X-ray emission unsettled. In the favored model, the emission is the sum of IC/IR flux and radiation from an additional component, likely a weak X-ray corona. The latter is needed to account for a fraction of the soft X-ray emission, while it should be sub-dominant at higher ($>10 \, \si{\kilo \electronvolt}$) energies. The deep \xmm{} observations, along with \nustar{} data, allowed us to refine the spectral parameters of the X-ray emission, with a photon index of $\Gamma \simeq 1.7$ and an intrinsic luminosity of $\mathcal{L}_{2-10 \, \si{\kilo \electronvolt}} \simeq 1.7 \times 10^{41} \, \si{\erg \, \second^{-1}}$. In \fig{\ref{fig: 1718-649 SED}}, we plot the observed and modeled SED of 1718-649 presented in \cite{Sobolewska2021}, adding the results of the joint-fit of \xmm{} and \nustar{} data. We also include the upper limit in the 20-40 keV measured from the \nustar{} observation. The improved X-ray dataset confirms the main findings of the SED modeling: while the IC/IR is the dominant contribution, it underpredicts the $< 1.5 \, \si{\kilo \electronvolt}$ flux by about a factor of 2. As proposed by \cite{Sobolewska2021}, this excess could be accounted for by an additional emitting component (modeled with a cutoff power law with $\Gamma = 2.0$, high energy cutoff set arbitrarily to 100 keV, and $\mathcal{L}_{2-10 \, \si{\kilo \electronvolt}} = 6.6\times10^{40} \, \si{\erg\, \second^{-1}}$) associated, i.e., with a weak X-ray corona or a radiatively inefficient nuclear emission \citep[e.g. an ADAF;][]{Ichimaru1977,Narayan1994,Narayan1995a,Narayan1995b,Abramowicz1995}. This model is consistent with the data at energies higher than 10 keV obtained from the \nustar{} observation which, unfortunately, provides only loose constraints beyond 20 keV. 

Finally, we also comment on the long-term X-ray variability of the source. \cite{Beuchert2018} found X-ray variability on timescale of years in the normalization of the power-law emission by a factor of about 2.5. In relation to the broadband model of \cite{Sobolewska2021}, this variability might be interpreted in terms of variations of seed infrared photons available for IC scattering, due to the presence of a clumpy circum-nuclear medium around the radio source \citep{Filippenko1985, Maccagni2014,Maccagni2018}. 

\subsection{1146+596 SED}
\begin{figure*}
     \centering
     \begin{subfigure}[]{0.49\textwidth}
         \centering
         \includegraphics[width=\textwidth]{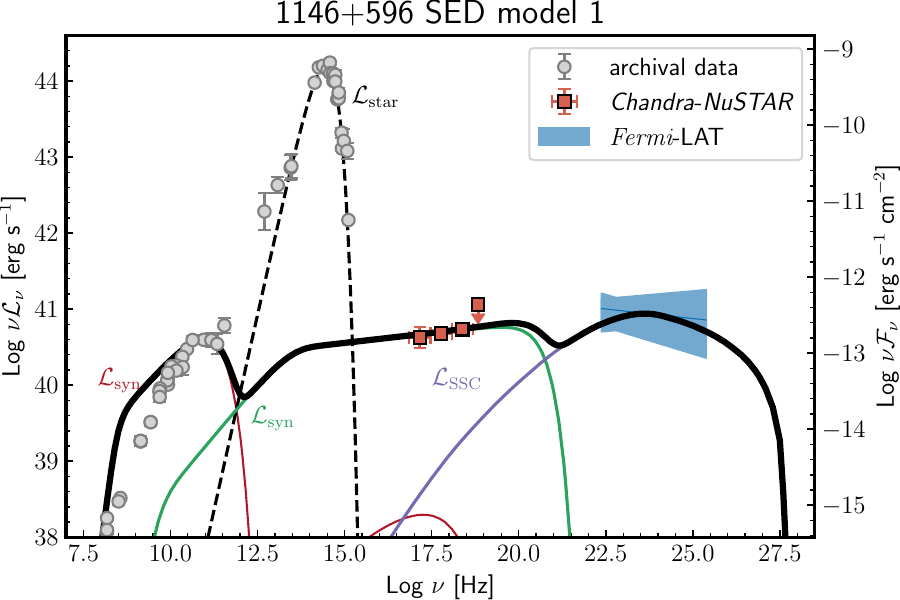}
         \caption{}
         \label{fig: 1146+596 SED model 1}
     \end{subfigure}
     \hfill
     \begin{subfigure}[]{0.49\textwidth}
         \centering
         \includegraphics[width=\textwidth]{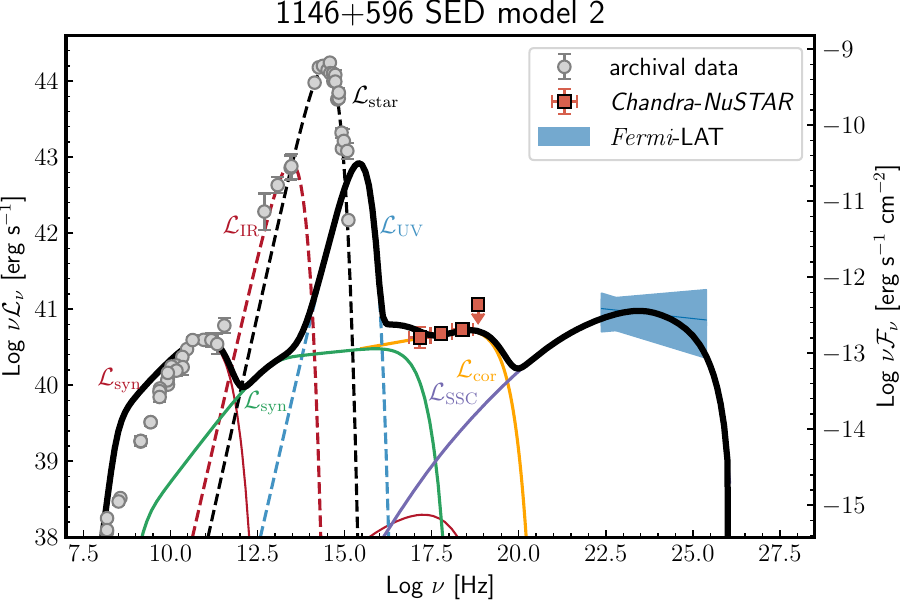}
         \caption{}
         \label{fig: 1146+596 SED model 2}
     \end{subfigure}
     \caption{Observed broadband spectral energy distribution of 1146+596 and theoretical model 1 (panel \ref{fig: 1146+596 SED model 1}) and model 2 (panel \ref{fig: 1146+596 SED model 2}). No formal fitting was performed. The model parameters are listed in Table \ref{tab: 1146+596 modeling}. Gray circles are data from \protect\cite{Balasubramaniam2021} and reference therein. Blue data are \fermi{} luminosity constraints from \protect\cite{Principe2020}. Dark orange data are \chandra–\nustar{} luminosity constraints from this work. Broadband model components are as follows: synchrotron and SSC curves in red, green, and violet \protect\citep[from][]{Massaro2006,Tramacere2009,Tramacere2011, Tramacere2020}, three black-body components representing the torus, the host galaxy, and the disk emission, in red, black, and blue, respectively. The solid orange line in model 2 illustrates the contribution of a low-luminosity X-ray nuclear emission (a weak X-ray corona or an ADAF-type emission). The black solid line is the sum of the different components related to the central AGN.}
     \label{fig: 1146+596 SED}
\end{figure*}

As mentioned in Section \ref{sec: introduction}, the X-ray continuum could be of non-thermal origin, related to the jets, or produced by the AGN powering the radio source. More insights can come from the study of the observed broadband SED. In the following, we explore the implications of a non-thermal origin of the high-energy emission of 1146+596 by modeling its radio-to-$\gamma$-ray emission. The radio-to-optical data are collected from the literature, the \fermi{} data are taken from \cite{Principe2020}. We added the X-ray data obtained from the broadband fit as part of this work. The observed SED is presented in \fig{\ref{fig: 1146+596 SED}}. 
The multi-wavelength data are not simultaneous; however, no variability has been observed over the timescales of years. 
The near-infrared to optical band is mainly contributed by the stellar continuum of the host galaxy. While we still display these data for completeness, they do not enter the modeling of the high-energy emission of the CSO.

In the multi-epoch 8.4 GHz VLBA maps, the source displays a pc-scale, two-sided structure, with mildly-relativistic, moderately-inclined jets \citep[][]{Principe2020}, embedded in a diffuse emission visible at lower frequencies \citep[][]{Taylor1998,Tremblay2016}.   
We thus test a leptonic model for the emission of the jets. 
We consider the simplest, and yet informative, scenario of a jet emitting via synchrotron mechanism and IC scattering.
Modeling of the SED was performed using the Jet SED modeler and fitting Tool \citep[\texttt{JetSeT},][]{Massaro2006,Tramacere2009,Tramacere2011,Tramacere2020}. In this model, the emission is produced in a jet region of spherical shape and radius $R$. The spherical volume is uniformly filled by relativistic plasma (i.e., a filling factor equal to one was assumed) and magnetic field, $B$. The energy densities of relativistic leptons and magnetic field are $\mathcal{U}_e$ and $\mathcal{U}_B$, respectively. The region is moving with a bulk Lorentz factor $\Gamma_\text{bulk}$, and $\theta$ is the angle between the jet axis and the observer line-of-sight (LOS). Interacting with the jet magnetic field, the electrons in the blob radiate via synchrotron emission. The synchrotron radiation provides the seed photons for the IC mechanism (SSC). A broken power-law function is used to describe the energy distribution of the radiating electrons, where $N \left(\gamma \right)$, $\gamma_\text{min}$, $\gamma_\text{max}$, and $\gamma_\text{break}$ are the minimum, maximum, and break Lorentz factors of the particles, and $p_1 = 2\alpha_1 + 1$ and $p_2 = 2\alpha_2 + 1$ are the spectral indices below and above $\gamma_\text{break}$, respectively.

We estimated the magnetic field within the lobes under equipartition conditions as in \cite{Wojtowicz2020}. Defining the effective radius of the lobes as $R_{\mathrm{eff}}= \sqrt[3]{\frac{3}{4}ab^2}$, where $a$ and $b$ are the major and minor axis of the CSO, respectively\footnote{Consequently, $LS=2a$.}, and assuming $b/a \simeq 0.25$ \citep{Kawakatu2008}\footnote{CSOs are expected to have more symmetric radio structures with respect to evolved FRII, for which $b / a \simeq 0.1$ are expected based on dynamical considerations.}, the equipartition magnetic field is
\begin{equation}
    B_{\mathrm{eq}} = \left[ 4.5 c_{12} \mathcal{L}_{\mathrm{rad}} \right]^{2/7} R_{\mathrm{eff}}^{-6/7} \, \si{\gauss} \simeq 10 \, \si{\milli \gauss}
    \label{eq: equipartition magnetic field}
\end{equation}

\noindent where $c_{12} \simeq 3 \times 10^7$ (in cgs units), and $\mathcal{L}_{\mathrm{rad}}$ stands for the total observed radio power in the frequency range 0.01–100 GHz \citep[see][]{Beck2005}. Here, we have assumed that the intrinsic radio continua could well be approximated by a simple power law with mean spectral index $\alpha=0.73$ above the peak frequency, as derived for young radio galaxies by \cite{deVries1997}. Under this assumption, 
\begin{equation}
    \mathcal{L}_{\mathrm{rad}} \simeq 7.62\times \mathcal{L}_{\mathrm{5 \, \si{\giga \hertz}}},
\end{equation}

\noindent where $\mathcal{L}_{5 \, \si{\giga \hertz}} = 1.62 \times 10^{23} \, \si{\watt\, \hertz^{-1}}$ \citep{Principe2020}.
The photon index of the $\gamma$-ray emission constrains the slope of the high-energy non-thermal emission, hence that of $N \left(\gamma\right)$ beyond the energy break, $p_2=2\Gamma-1$. 
Indeed, these are initial estimates which need to be adjusted for the model to match the observed SED.

As a first test, we tried to model the radio, X-ray, and $\gamma$-ray emission in the framework of a synchrotron and SSC, one-zone scenario. However, this model has issues in properly reproducing the X-ray emission regardless of whether it is ascribed to synchrotron or SSC radiation. Therefore, we relaxed the hypothesis of a single emitting region and assumed that the dominant contribution at low/radio frequencies is from a larger region, either in the jets or forming lobes, while the high-energy emission is produced in a smaller/inner jet region. The latter has a synchrotron curve extending to the UV-X-ray band, while the SSC component is responsible for the $\gamma$-ray emission. This model (Model 1) is borrowed from high-energy peaked BL Lacs and is also motivated by the fact that, in the $\gamma$-ray luminosity \emph{vs.} photon index plot \citep[see \fig 7 in][]{Principe2020}, 1146+596 is located close to the region of $\gamma$-ray BL Lacs. Hence, in this scenario, 1146+596 could be considered as a misaligned, low-power BL Lac with a mildly relativistic jet. 

\begin{figure}
    \centering
    \includegraphics[width=0.48\textwidth]{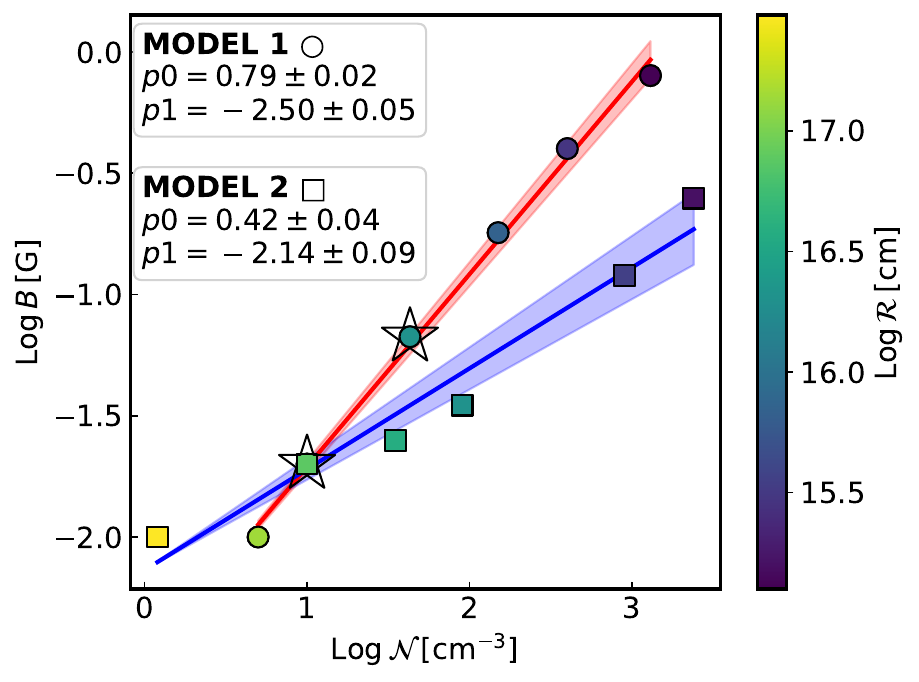}
    \caption{Parameter space of SED models for 1146+596. Data points are shown with circles and squares for Model 1 and 2, respectively. The linear relations between the magnetic field strength and the particles density found for Models 1 and 2 are reported in red and blue, where $p0$ and $p1$ are the angular coefficient and the normalization.
    A star marker is used to highlight models for which parameters are listed in \tab{\ref{tab: 1146+596 modeling}}.}
    \label{fig: 1146+596 SED params}
\end{figure}

Model 1 is presented in \fig{\ref{fig: 1146+596 SED model 1}} and its parameters are reported in \tab \ref{tab: 1146+596 modeling}. We started fixing all parameters, excepting for $R$, $B$, and $\mathcal{N}$. Furthermore, we set the lower limit for the magnetic field equal to the equipartition magnetic field $B_{\mathrm{eq}}\simeq 10 \, \si{\milli \gauss}$. This choice is driven by the assumption that the magnetic field strength is expected to decrease along the jet axis \citep{OSullivan2009,Pudritz2012,Gabuzda2017,Massaglia2019}. Since no strong shocks are observed at the end of the jet, we consider this value as the lowest value for the magnetic field strength. Other limits in the parameter space are set by the size of emitting blob: on the one hand, we assumed a lower limit for the radius of the emitting region of the order of $10^{15} \si{\centi \meter}$, in agreement with the typical values found for blazars \citep[e.g.][]{Celotti2008}; on the other hand, we set the size of the radio source as upper limit ($\sim 1 \, \si{\parsec}$). With these constraints, we modeled the SED of 1146+596. We note that there is not a unique set of parameters' values which provide a satisfactory modeling. This type of degeneracy in non-thermal modeling of jetted AGN is something well known and established \citep[e.g.][]{Bottcher2007,Lucchini2019}. For this reason, we investigated the parameters space in further details. In particular, we found a linear relation between the magnetic field strength and the radiating particles density in bi-logarithmic scale: we argue that all models located on this relation shown in red in \fig{\ref{fig: 1146+596 SED params}} well describe the observed SED.
Similarly to what is observed in BL Lacs, this scenario implies a strongly particle-dominated inner-jet emitting region, with $\mathcal{U}_B/\mathcal{U}_e$ ranges from 1$e$-3 up to 3$e$-2. Moreover, in order to produce the observed X-ray emission via synchrotron mechanism, a $\log_{10} \gamma_\text{max} \geq 7.5$ is needed: this is an extreme value even for BL Lacs detected at TeV energies \citep{Celotti2008}. However, theoretical works and simulations \citep[e.g.][]{Bottcher2010,Mimica2012,Vaidya2018,Mukherjee2021} have suggested that internal-shocks can efficiently accelerate particles up to such high values of $\gamma_{\max}$, lending support to a synchrotron origin of the X-ray emission \citep[e.g.,][and references therein]{Bottcher2010}. This means that, if our model 1 is correct, extremely efficient acceleration processes are in action in 1146+596.
Energy equipartition between particles and $B$ field is instead assumed for the region producing the bulk of the radio emission. At high energies, the contribution of this component is negligible, and, interestingly, it overestimates the observed radio fluxes in the range $\sim 0.3 - 3 \, \si{\giga \hertz}$ despite of including the synchrotron-self absorption in the model. One possibility is that an additional effect is shaping the synchrotron emission and this could be free-free absorption by clouds nearby the center of the host galaxy, as for 1718-649 \citep{Tingay2003}. 

Assuming Model 1, we can explain very well the observed X-emission, as well as $\gamma$-ray emission, via synchrotron and SSC radiation, respectively. However, as discussed above, very high values of $\gamma_{\max}$ and, consequently efficient acceleration processes, are required. Hence, we investigated an alternative scenario and relaxed the hypothesis of the X-ray flux being produced via synchrotron emission. As discussed in Section \ref{sec: 1146+596}, 1146+596 harbors a LINER type AGN as 1718-649. Therefore, the AGN emission could contribute to the X-ray band. We assumed a phenomemenological AGN model similar to the one of adopted for 1718-649 in \cite{Sobolewska2021} (see Section \ref{sec: SED 1718-649}): a disk black-body peaking at UV frequencies and a cut-off power-law function with $\Gamma = 1.9$, $E_\text{cut-off} = 100 \, \si{\kilo \electronvolt}$. We adjusted the luminosities of these two components to match the observed SED: this resulted in a UV luminosity of $\mathcal{L}_\text{UV} \simeq 8 \times 10^{42} \, \si{\erg \, \second^{-1}}$, which is in agreement with the upper limit of $\mathcal{L}_\text{bol} < 5 \times 10^{43} \, \si{\erg \, \second^{-1}}$ reported in \cite{Balasubramaniam2021}, and a $2- 10 \, \si{\kilo \electronvolt}$ luminosity of $\mathcal{L}_{2-10 \, \si{\kilo \electronvolt}} \simeq  9 \times 10^{40} \, \si{\erg \, \second^{-1}}$. Model 2 is shown in \fig{\ref{fig: 1146+596 SED model 2}} and the model parameters are reported in \tab \ref{tab: 1146+596 modeling}. Adopting these parameters, typical for LINER-type AGN \citep[see ][]{Gonzalez2009}, the X-ray emission could be entirely explained as emission from a weak corona or a radiatively inefficient accretion flow \citep{Ichimaru1977,Narayan1994,Narayan1995a,Narayan1995b,Abramowicz1995,Chen1995}. The SSC emission of the inner-jet emerges in the $\gamma$-ray band, while its synchrotron radiation is hidden below the thermal emission of the host galaxy and of the AGN (for completeness, we also included the emission of the putative torus, modeled as a black-body peaking in the IR band).
Following the same approach adopted for Model 1, we investigated the degeneracy in the parameters space. We found that all models lying on the relation in blue in \fig{\ref{fig: 1146+596 SED params}} provide a good modeling of the observed SED. In these models, the X-ray emission is now ascribed to the AGN. 
The dominance of the particles over the magnetic field holds true also in this case, with $\mathcal{U}_B/\mathcal{U}_e$ ranging from $8 \times 10^{-4}$ up to $5 \times 10^{-3}$. 

It has to be noticed that Model 2 is less constrained than Model 1, as arises from \fig{\ref{fig: 1146+596 SED params}}: if on one hand the synchrotron component is well constrained by the observed X-ray emission in Model 1, in Model 2 we can only set upper limits on it since the main contribution to the observed X-ray emission comes from the additional component.

\subsubsection{Jet power}
Estimates of the total power carried by the jet can be inferred from the SED modeling. The jet power ($\mathcal{L}_\text{jet}$) is the sum of the kinetic power ($\mathcal{L}_\text{kin}$), which is the energy stored into relativistic emitting electrons ($\mathcal{L}_e$) and cold protons ($\mathcal{L}_p$), and the power carried as Poynting flux ($\mathcal{L}_B$). Here, we assumed one cold proton per radiating electron.
A fraction of this power is radiatively dissipated, the radiative power ($\mathcal{L}_\text{rad}$). These quantities have been estimated with the formulae in \cite{Zdziarski2014} and the values for the two models are reported in \tab{\ref{tab: 1146+596 modeling}}. From the study of the space parameters in both models, we can put constraints on the energetic budget of the source. The total jet powers are in the range 
$0.07 - 4 \times 10^{42}  \, \si{\erg{} \, \second^{-1}}$ for Model 1 and $0.1 - 2 \times 10^{42}  \, \si{\erg{} \, \second^{-1}}$ for Model 2. 

For both models, the Poynting flux is a minor contribution to the total power, as expected being $\mathcal{U}_e \gg \mathcal{U}_B$ in both cases. The bulk of the jet power lies in the kinetic energy of the radiative electrons, while the hadronic component contributes with few percentages to the total kinetic energy of the jet ($\sim 20\%$ and $\sim 10\%$ for Model 1 and 2, respectively). 
The jet radiates away a few percentages of its total power ($\lesssim 10\%$), meaning that the bulk of $\mathcal{L}_\text{jet}$ can be used to expand the radio structures.

\begin{table*}
    \centering
    \caption{1146+596 input SED models parameters and estimated jet powers.}
    \begin{tabular}{lccccc}
    \hline
    \hline
    & & & \multicolumn{2}{c}{Jet} & Mini lobe\\
    \hline
    Description & Symbol & Unit & Model 1 & Model 2 & Model\\
    (1) & (2) & (3) & (4) & (5) & (6)\\
    \hline \\[-2.5mm] 
    \multicolumn{6}{c}{Input Parameters} \\ \\[-2.5mm]
    \hline
    Radius of the emitting region & $\log_{10} R$ & $\si{\centi \meter}$ & 16.30 & 16.88 & 18.74\\
    Lorentz factor & $\Gamma$ & & 1.11 & 1.04 & 1.00\\
    Jet viewing angle & $\theta$ & degree & 10 & 21 & 21 \\
    Magnetic field & $B$ & $\si{\milli \gauss}$ & 67 & 20 & 10\\
    Emitting electron number density & $\log_{10} \mathcal{N}$ & $\si{\centi \meter^{-3}}$ & 1.63 & 1 & -1.05\\
    Low-energy slope & $p_1$ & & 1.9 & 1.8 & 2\\
    High-energy slope & $p_2$ & & 2.9 & 2.9 & ---\\
    Lorentz factor (min) & $\log_{10} \gamma_{\min}$ & & 2 & 2 & 1\\
    Lorentz factor (break) &  $\log_{10} \gamma_{\text{break}}$ & & 4 & 4 & ---\\
    Lorentz factor (max) & $\log_{10} \gamma_{\max}$ & & 7.5 & 6 & 3.4\\
    Energy density ratio of magnetic field and electrons & $\mathcal{U}_B/\mathcal{U}_e$ & & 3$e$-2 & 2$e$-3 & 1\\ 
    \hline \\[-2.5mm]
    \multicolumn{6}{c}{Additional Power-law X-Ray Component} \\ \\[-2.5mm]
    \hline
    Photon index & $\Gamma$ & & --- & 1.9 & ---\\
    Luminosity & $\mathcal{L}_{\mathrm{2-10\, keV}}$ & $\si{\erg \, \second^{-1}}/10^{41}$ & --- & 8$e$-1 & ---\\
    \hline \\[-2.5mm]
    \multicolumn{6}{c}{Jet Power} \\ \\[-2.5mm]
    \hline
    Radiative power & $\mathcal{L}_{\text{rad}}$ & $\si{\erg \, \second^{-1}}/10^{41}$ & 8$e$-1 & 1.1 & 4$e$-1\\
    Electrons kinetic power & $\mathcal{L}_{e}$ & $\si{\erg \, \second^{-1}}/10^{42}$ & 5$e$-1 & 1.1 & 6{\it e}-2\\
    Protons kinetic power & $\mathcal{L}_{p}$ & $\si{\erg \, \second^{-1}}/10^{41} $ & 1.3 & 9$e$-1 & 2$e$-4\\
    Total kinetic power & $\mathcal{L}_{\text{kin}}$ & $\si{\erg \, \second^{-1}}/10^{42} $ & 6$e$-1 & 1.2 & 6$e$-2\\
    Magnetic power & $\mathcal{L}_{B}$ & $\si{\erg \, \second^{-1}}/10^{41}$ & 1$e$-1 & 7$e$-2 & 1.3\\
    Total jet power & $\mathcal{L}_{\text{jet}}$ & $\si{\erg \, \second^{-1}}/10^{42}$ & 6$e$-1 & 1.2 & 2$e$-1\\
    \hline
    \end{tabular}
    \tablefoot{
    (1) Description of the input parameter, (2) symbol used for the input parameter, (3) unit in which the parameter is expressed, (4-5) parameters values for the jet models, (6) parameters values for the mini-lobe model. Powers are estimated with formulae in \cite{Zdziarski2014}, that are accurate for small bulk motions.}
    \label{tab: 1146+596 modeling}
\end{table*}

\section{Discussion}
\label{sec: discussion}

\subsection{Ambient medium of CSO}
\label{sec: ambient medium}

\begin{figure}
    \centering
    \includegraphics[width=0.48\textwidth]{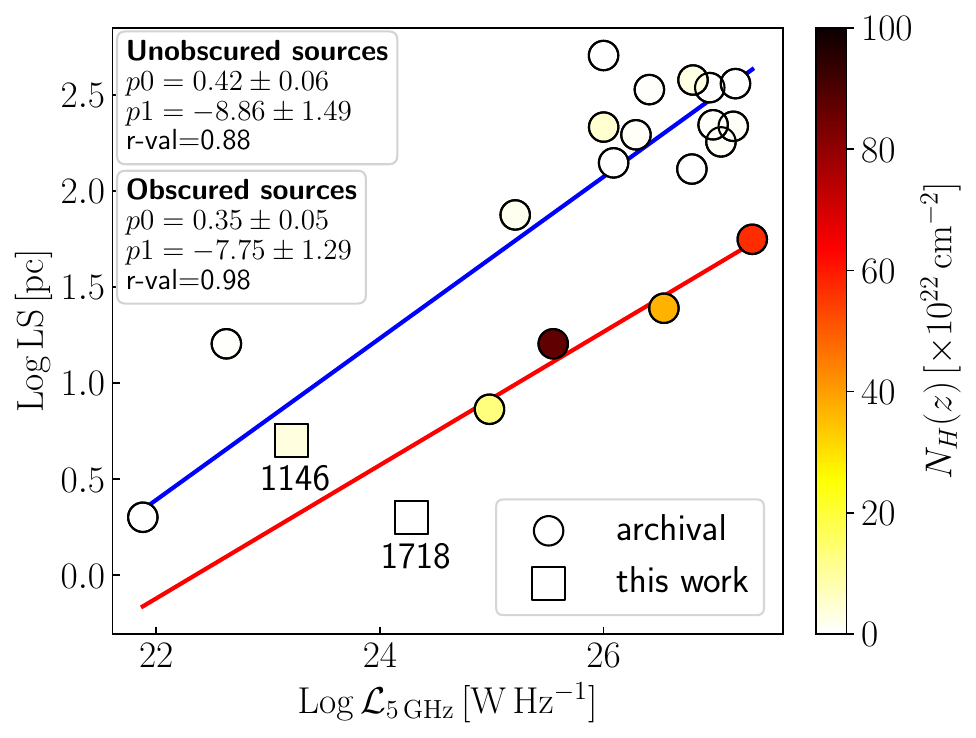}
    \caption{Luminosity at 5 GHz vs. radio source size proposed by \protect\cite{Sobolewska2019b} (S19). Literature data are selected as in Section \ref{sec: ambient medium}. 1718-649 and 1146+596 are marked with squares. Color coding indicates the intrinsic equivalent hydrogen column density measured from the X-ray spectra. Red and blue solid lines mark the linear relations fitted on obscured ($N_{\rm H} \geq 10^{23} \, \si{\centi \meter^{-2}}$) and unobscured ($N_{\rm H} < 10^{23} \, \si{\centi \meter^{-2}}$) sources, where $p0$ and $p1$ are the angular coefficient and the normalization, respectively. The r-val stands for the Pearson correlation coefficient. Adapted from S19.}
    \label{fig: radio lum vs size}
\end{figure}

X-rays have a potential to uncover the role of ambient medium in the source expansion: the host galaxy environment can be studied in the X-ray band via the detection of the hot component of the ISM and the total equivalent hydrogen column density along the LOS. S19 noted that obscured and unobscured CSO sources appeared to occupy different regions in the linear size ($LS$) vs. radio power ($\mathcal{L}_{5 \, \si{\giga \hertz}}$) plane, possibly indicating a role of the ambient medium in the source expansion.
The two sources presented in this work have low-to-moderate levels of obscuration, with intrinsic equivalent hydrogen column densities in the range $\sim10^{21} \, \si{\centi \meter^{-2}}$ to a few $10^{22} \, \si{\centi \meter^{-2}}$, in agreement with the majority of young radio sources reported by S19. In \fig{\ref{fig: radio lum vs size}}, we updated the $LS$-$\mathcal{L}_{5 \, \si{\giga \hertz}}$ plot in S19, with the new $N_{\rm H}$ estimates in \cite{Sobolewska2023} and in this work. In addition, we selected sources belonging to the sample of {\it bona-fide} CSOs built by \cite{Kiehlmann2023} to avoid contamination. We excluded 4 sources from our analysis: NGC 7674, 4C+37.11, 4C +52.37, J1247+6723. The first one exhibits peculiar properties in terms of linear size, and it is considered an outlier also in \cite{Kiehlmann2023}, while we could not find estimates of the intrinsic obscuration in the literature for the other three sources.

\cite{Orienti2014} pointed out a positive correlation between radio luminosity and radio linear size in CSOs. Hence, we investigate this correlation splitting sources in two classes: obscured, with $N_{\rm H} \geq 10^{23} \, \si{\centi \meter^{-2}}$, and unobscured, with $N_{\rm H} < 10^{23} \, \si{\centi \meter^{-2}}$. For both samples, the Pearson coefficients (r-val) confirm a strong positive correlation for both samples of obscured (r-val=0.98) and unobscured (r-val=0.88) CSOs. Furthermore, in order to estimate the statistical significance of this separation in obscured and unobscured CSOs, we performed two 1-dimensional 2-sample Kolmogorov-Smirnov (K-S) tests concerning the CSO radio luminosity and radio linear size. We found that while the radio luminosities of the two samples are likely to come from the same distribution ($p$-value = 0.96), we cannot reject the hypothesis that the radio sizes of the two samples come from two distinct distributions ($p$-value = 0.02). Even if the selected samples over which the test was performed slightly differs, due to the different selection {\emph critieria}, our results are still in a good agreement with \cite{Sobolewska2019b}, who found $p$-value = 0.8 and $p$-value = $9.8 \times 10^{-4}$ for radio luminosity and radio size, respectively.
From \fig{\ref{fig: radio lum vs size}}, we note that 1146+596 lies, as expected, on the branch of unobscured/mildly obscured sources, while 1718-649 seems instead to share the same location as the obscured sources, contrary to what is expected from their moderate/low column densities. The reason for the location of 1718-649 in the $LS$-$\mathcal{L}_{5 \, \si{\giga \hertz}}$ diagram remains obscure. A possibility is that, for unclear reasons, 1718-649 exhibits significantly higher luminosity at 5 GHz than its counterparts at the same linear size, or it has a more compact radio structure compared to others with similar radio luminosities. Alternatively, the separation in obscured and unobscured sources is strongly biased by the small available statistics, and we are looking at only a small fraction of CSOs that is not representative of the entire population. Clearly, this represents a limit for our statistical tests and prevents us from strong conclusions. Moreover, it should also be taken into account that the X-ray analysis cannot constrain the location of the obscuring component precisely because of the limited angular resolution of the X-ray instruments ($\sim 0.5 \si{\arcsecond}$ with \chandra). As a consequence, it is possible that the X-ray obscurer is also distributed at the same scales, thus obscuring the X-ray emission without affecting the CSO’s expansion. More insights on the spatial distribution of the obscuring material can come from $HI$ measurements. In general, measurements of $HI$ absorption features in CSOs pointed out column densities 1-2 order of magnitudes lower than those estimated by $N_{\rm H}$ in X-rays. Indeed, it is important to bear in mind that $N_{\rm H}$ quantifies the column density across all gas phases, while $N_{\rm HI}$ only traces the column density through the atomic hydrogen\footnote{Note that the $N_{\rm HI}$ value is directly proportional to the Spin Temperature, and an underestimation of this parameter could potentially lead to a reduced $N_{\rm HI}$ value.}. However, correlations between $N_{\rm H}$ in X-rays and $N_{\rm HI}$ in radio bands have been found for CSOs, even if with large scatters \citep{Ostorero2010,Ostorero2017}. \cite{Ostorero2017} found that
\begin{equation}
    \mathrm{Log} \, N_{\rm HI} = 13.5_{-8.4}^{+5.8} + 0.32_{-0.27}^{+0.38} \times \mathrm{Log} \, N_{\rm H},
\end{equation}
with an intrinsic spread of $\sigma_{\mathrm{int}} = 0.59_{-0.18}^{+0.38}$ \citep[see][for further details]{Ostorero2017}. This value of $\sigma_{\mathrm{int}}$ implies that the corresponding $N_{\rm HI}$ falls is expected to be a factor of $\sim 4$ from the mean relation at $1 \, \sigma$. With our estimates of the intrinsic $N_{\rm H}$ of 1146+596 ($N_{\rm H} \simeq 3.5 \times 10^{22} \, \si{\centi \meter^{-2}}$), we expect $N_{\rm HI}$ to be within a factor of 4 from $\sim 5 \times 10^{20} \, \si{\centi \meter^{-2}}$. This value is perfectly in agreement with measurements in \cite{Peck1998}, who found $N_{\rm HI}=2-12 \times 10^{20} \, \si{\centi \meter^{-2}}$, suggesting that the $N_{\rm H}$ measured in X-ray is likely located nearby the CSO. Hence, in the specific case of 1146+596, the presence of a dust lane along the major axis of the host galaxy observed with the optical Hubble Space Telescope \citep{Perlman2001} does not contribute to the estimated $N_{\rm H}$ significantly. Similar conclusions were reached also for 1718-649 \citep{Ostorero2017}, suggesting that little amount of gas ($N_{\rm H} \simeq 9 \times 10^{20} \, \si{\centi \meter^{-2}}$) should be located nearby the central CSO. By the simplest assumption of a uniform distribution of the absorbing column over a specific surface area, we can infer a rough estimate of the amount of gas necessary to produce the derived X-ray obscuration in our sources. The total amount of gas is given by
\begin{equation}
    \mathcal{M}_{\mathrm{gas}} \simeq N_{\rm H} \pi r^2 m_p
    \simeq 2.5 \times 10^8 \mathcal{M}_{\odot} \left(\frac{N_{\rm H}}{10^{22} \, \si{\centi \meter^{-2}}} \right) \left(\frac{r}{1 \, \si{\kilo \parsec}} \right)^2,
\end{equation}
where $N_{\rm H}$ is the equivalent hydrogen column density derived from X-ray observations, $r$ is the sphere radius over which the gas is distributed, and $m_p$ is the proton mass. Assuming $r$ of the order of typical radii of the tori in AGN \citep[$r\simeq 10-100 \, \si{\parsec}$, e.g.,][]{Almeida2017}, we get $\mathcal{M}_{\mathrm{gas}} \simeq 10^3-10^5 \mathcal{M}_{\odot}$ and $\mathcal{M}_{\mathrm{gas}} \simeq 10^5-10^7 \mathcal{M}_{\odot}$ for 1718-649 and 1146+596, respectively.

In both sources, 1718-649 and 1146+596, thermal emission is evident in the soft X-ray band ($<2 \, \si{\kilo \electronvolt}$). The emission is extended on kiloparsec scales, thus beyond the size of the CSOs, raising the question of the origin of such emission and its relation with the presence of a young source. Our results on 1718-649 agree with the analysis performed by \cite{Beuchert2018}. The authors associated this emission with supernovae explosions in the host galaxy. The joint analysis of \chandra{} and \nustar{} data of 1146+596 suggested the presence of a multi-temperature gas ($kT \simeq 0.3 \, \si{\kilo \electronvolt}$ and $kT \simeq 1.2 \, \si{\kilo \electronvolt}$). The warmer temperature lies on the relation $L_X \left( \text{gas}\right)$-$kT \left( \text{gas}\right)$ for the hot ISM typically found in early-type galaxies \citep{Fabbiano2019}, as the host of 1146+596. The nature of the hotter component is instead more elusive. It could be emitted by gas either photo-ionized by the central AGN or heated by the shocks driven by the impact of the radio jet on the ISM clouds, or even a superposition of these two processes \citep[e.g.,][and references therein]{Fabbiano2019}.  Even if, thus far, signatures of jet-ISM interactions have been typically observed in more evolved radio sources, like CSS with linear sizes from 500 pc up to 20 kpc \citep[e.g.,][and references therein]{ODea2021}, evidence in favor of such interactions have been reported also in CSOs \citep[e.g.][and references therein]{Orienti2016}. 
In the case of 1146+596, \cite{Taylor1998} report kpc-scale radio halo ($LS \sim 1 \, \si{\kilo \parsec}$), which could be a relic of a previous radio cycle. A shock-heated gas could be then related to recurrent radio outbursts. 
This scenario is also supported by numerical simulations \citep{Zovaro2019b,Zovaro2019a}, which showed that outflows of CSOs can reach scales several times larger than the radio source size. As a consequence, a relation between the properties observed on host galaxy scales and the central source cannot be excluded a \emph{priori}.
Alternatively, since the \chandra{} inspection revealed the presence of few individual sources ($\sim 4$), that we excluded from our analysis, a contamination from a small population of individual sources, such as Low-Mass X-ray Binaries or supernovae, nearby the central AGN is an alternative possibility \citep[e.g.][]{Lehmer2016}. In this scenario, the hotter gas has no direct connection with the central source.

\subsection{Origin of high-energy emission}

\cite{Sobolewska2021} adopted the dynamical and radiative model by \cite{Stawarz2008} to explain the broadband SED of 1718-649. In this model, both a weak corona and IC scattering of infrared photons by the lobes’ electrons contribute to the observed X-ray emission, while the GeV emission is produced via IC scattering of the UV photon field by the lobes’ electrons. This scenario is supported by the lobe-dominated radio structure of 1718-649, pointing to large viewing angles and possible de-boosting of the jet emission. The radio morphology of 1146+596, instead, appears to be jet-dominated \citep[see Fig. 4 in][]{Principe2020}. Thus, we adopted a different model for 1146+596, in which the $\gamma$-ray emission is produced in a blob within the jet. In both versions (Models 1 and 2), the GeV emission is produced via IC scattering on synchrotron photons, while the origin of the X-ray emission differs. In the former, the X-rays are produced via synchrotron radiation, and the source appears similar to a high-energy peaked BL Lacs; in the latter, the X-ray come from a weak corona or an ADAF-type disk, hence they are produced by the central AGN. Note that, while a corona contribution to the X-ray emission, although not dominant, is included also in the SED modeling of 1718-649 \citep{Sobolewska2021}, a scenario of X-ray synchrotron emission, as our Model 1, would not be easy to accommodate given the harder X-ray photon index of 1718-649.

Interestingly, \cite{Mukherjee2017} argued that jets with powers $\lesssim 10^{43} \, \si{\erg \, \second^{-1}}$, such as those of 1718-649, may be too weak to escape the ISM confinement, and too weakly pressurized to prevent an infall of gas back into the initially created central cavity. 
Noteworthy, our estimates suggest that also 1146+596 is characterized by a jet power $\lesssim 10^{42} \, \si{\erg \, \second^{-1}}$. Moreover, it has to be noted that our estimates of the jet power of 1146+596 have to be interpreted as an upper limit. Indeed, no external Compton contributions have been taken into account to model the high-energy emission of the source, contrary to what has been done with 1718-649. We note that this addition would return a lower estimate of the jet power. 
For the accretion disk luminosity, $\mathcal{L}_\text{disk} \simeq 5 \times 10^{43} \, \si{\erg \, \second^{-1}}$, estimated by \cite{Balasubramaniam2021} based on the $8-1000 \, \si{\micro \meter}$ fluxes in \cite{Kosmaczewski2020}, the jet to disk luminosity ratio is in the range $\lesssim 0.1$. These values are in agreement with the results by \cite{Ostorero2010} for a sample of CSOs, who found $\mathcal{L}_\text{jet}/\mathcal{L}_\text{disk} = 0.01-0.1$, while there are indications of higher ratios, $\sim 1 - 10$, in powerful blazars \citep{Celotti2008}.

\section{Conclusions}

In this work, we analyzed archival and new X-ray observations of two confirmed $\gamma$-ray emitters CSOs: 1718-649 and 1146+596. In particular, for the first time, we report on their detections with \nustar{} at energies $> 10 \, \si{ keV}$. Our main findings are as follows:
\begin{itemize}
    \item the joint \xmm{} and \nustar{} observations of 1718-649, extending the X-ray energy range up to $19 \, \si{\kilo \electronvolt}$, confirmed the results of previous works \citep{Siemiginowska2016,Beuchert2018}: the primary X-ray emission is modeled with a mildly absorbed ($N_{\rm H} = 0.09_{-0.01}^{+0.01} \times 10^{22} \, \si{\centi \meter^{-2}}$) power-law model ($\Gamma = 1.76_{-0.05}^{+0.05}$) and a diffuse thermal X-ray emission ($kT = 0.72_{-0.06}^{+0.06} \, \si{\kilo \electronvolt}$);
    \item the joint fit of the \chandra{} and \nustar{} spectra of 1146+596 indicates a softer photon index ($\Gamma = 1.92_{-0.33}^{+0.34}$) with respect to the values in literature, with a moderate intrinsic obscuration ($N_{\rm H} = 3.49_{-1.04}^{+1.28} \times 10^{22} \, \si{\centi \meter^{-2}}$). From the fit, we estimated an intrinsic X-ray luminosity of $\mathcal{L}_{2-10 \, \si{\kilo \electronvolt}} = \left(6.0 \pm0.4\right) \times 10^{40} \, \si{\erg \, \second^{-1}}$;
    \item we identify a multi-temperature thermal component ($kT = 0.32_{-0.09}^{+0.17} \, \si{\kilo \electronvolt}$, $kT = 1.16_{-0.18}^{+0.31} \, \si{\kilo \electronvolt}$) in the X-ray spectrum of 1146+596. The thermal emission is extended beyond the radio size of the CSO. The cooler component is likely related to the emission of the hot ISM of the host galaxy, while the higher-temperature one could be the signature of jet-ISM interactions; 
    \item we investigated the role of the ambient medium in confining the source expansion through the study of intrinsic obscuration in X-rays. Based on the location of our sources in the diagnostic {\it LS vs.} $\mathcal{L}_{5 \, \si{\giga \hertz}}$ plot proposed in S19, 
    we argue that heavily obscured, and possibly frustrated sources, show different radio sizes with respect to unobscured, free to expand, ones. This separation is supported by a 1-dimensional 2-sample Kolmogorov-Smirnov ($p$-value = 0.02).
    \item we modeled, for the first time, the broadband SED of 1146+596 in the framework of a leptonic synchrotron-SSC model and proposed two scenarios. In the former, the X-ray emission is due to synchrotron emission: 1146+596 can be seen as a low-power, moderately misaligned BL-Lac. In the latter, the X-ray emission is related to the AGN, either to a weak corona or an ADAF, similarly to the model proposed for 1718-649 by \cite{Sobolewska2021}. The maximum estimated jet is, for both models, $\sim 10^{42} \, \si{\erg \, \second^{-1}}$, and places 1146+596 among the low-power young radio sources, which might not expand beyond the host galaxy limits.
\end{itemize}

The detection in $\gamma$-rays of CSOs is a clear confirmation of a non-thermal high-energy component. However our work corroborates the idea of different sites for the origin of such emission (in the lobes or the jets) and of an additional contribution (a weak corona or an ADAF-like emission), which is present, possibly dominant, in the X-ray band. We note that, depending on the site of production, the non-thermal high-energy emission will change differently as the source evolves: the isotropic lobe emission is expected to fade, while the jet-related one should persist (unless of a change of the jet parameters). A test of this scenario requires sensitive high-energy observations of samples of radio sources in the first stages of their evolution, which unfortunately are not yet available \citep[see e.g.][]{Principe2021}. Given the source’s radio morphology and constraints on the dynamics, we modeled high-energy emission of 1146+596 with a simple, and yet informative, SSC model. However, as in the case of 1718-649, EC against, e.g., photons coming from the disk and/or the dusty torus might contribute to the high-energy emission observed in 1146+596. In general, the EC contribution at high-energies would lead to a lower estimate of the total jet power of the source.

A larger number of obscured ($N_{\mathrm{H}} > 10^{23} \, \si{\centi \meter^{-2}}$) CSOs is also needed to investigate the role of the surrounding ambient medium in confining the source expansion. Combining this with detailed studies of the spatial distribution, composition (both in atomic and molecular phase) and kinematics of the gas close to the nuclear regions, thanks to the current radio and mm facilities, can provide essential insights into AGN feedback processes in CSOs.

\begin{acknowledgements}
      Authors thank the anonymous referee for useful suggestions that improved the clarity of the manuscript. Authors also thank E. Torresi for the useful discussions on X-ray modeling of thermal and photoionized gasses. M.S. and A.S. were supported by NASA contract NAS8-03060 (Chandra X-ray Center). M.S. acknowledges NASA/Chandra contract GO2-23110X. {\L }.S. was supported by the Polish NSC grant 2016/22/E/ST9/00061. G.P. acknowledges the support by ICSC – Centro Nazionale di Ricerca in High Performance Computing, Big Data and Quantum Computing, funded by European Union – NextGenerationEU.
\end{acknowledgements}

\bibliographystyle{aa} 
\bibliography{bibliography}
\end{document}